\documentclass[preprintnumbers, onecolumn, floatfix, preprintnumbers, amsmath, amssymb, superscriptaddress]{revtex4}

\usepackage[colorlinks,linkcolor=red,urlcolor=blue,citecolor=blue]{hyperref}
\usepackage{amssymb, amsmath, bm, dcolumn, epsf, graphicx, latexsym, mathbbol, slashed}
\usepackage{mathrsfs}
\usepackage{comment}

\begin{document}

\title{Gravitational Waveform and Polarization
from Binary Black Hole Inspiral in Dynamical Chern-Simons Gravity: From Generation to Propagation}

\author{Zhao Li}
\email{lz111301@mail.ustc.edu.cn}
\affiliation{ Department of Astronomy, University of Science and Technology of China, Hefei, Anhui 230026, China,\\ and School of Astronomy and Space Science, University of Science and Technology of China, Hefei 230026, China}

\author{Jin Qiao}
%\email{qiaojin@zjut.edu.cn}
\affiliation{Purple Mountain Observatory, Chinese Academy of Sciences,
Nanjing, 210023, China,\\ and School of Astronomy and Space Science, University of Science and Technology of China, Hefei 230026, China}

\author{Tan Liu}
%\email{lewton@hust.edu.cn}
\affiliation{ MOE Key Laboratory of Fundamental Physical Quantities Measurements, Hubei Key Laboratory of Gravitation and Quantum Physics, PGMF, Department of Astronomy, and School of Physics, Huazhong University of Science and Technology, Wuhan 430074, China}

\author{Tao Zhu}
\affiliation{ Institute for theoretical physics and cosmology, Zhejiang University of Technology, Hangzhou, 310032, China,
\\ United Center for Gravitational Wave Physics (UCGWP),
Zhejiang University of Technology, Hangzhou, 310032, China}

\author{Wen Zhao}
\email{wzhao7@ustc.edu.cn}
\affiliation{ Department of Astronomy, University of Science and Technology of China, Hefei, Anhui 230026, China,\\ and School of Astronomy and Space Science, University of Science and Technology of China, Hefei 230026, China}

\begin{abstract}
We calculate the gravitational waveform radiated from spinning black holes (BHs) binary in dynamical Chern-Simons (dCS) gravity. The equation of motion (EOM) of the spinning binary BHs is derived based on the modified Mathisson-Papapetrou-Dixon equation for the spin-aligned circular orbits. The leading-order effects induced by the dCS theory contain spin-spin interaction and monopole-quadrupole interaction, which influences both the EOM of the binary system and corresponding gravitational waveform at the second post-Newtonian (PN) order (i.e., 2PN order). After reporting the waveforms, we investigate the polarization modes of gravitational waves (GWs) in dCS theory. None of the extra modes appears in this theory up to the considered PN order. Moreover, since the time scale of the binary merger is much smaller than that of the cosmological expansion, the parity-violating effect of the dCS theory does not appear in the process of GW generation. However, during the process of GW propagation, amplitude birefringence, a typical parity-violating effect, makes plus and cross modes convert to each other, which modifies the gravitational waveform at 1.5PN order.
\end{abstract}

\maketitle

\section{\label{Introduction}Introduction}
General relativity (GR) is always considered as the most successful theory of gravity. However, various difficulties of this theory are also well known. For instance, in the theoretical side, GR has the singularity and quantization problems \cite{Kiefer2007}. In the experimental side, all the observations in cosmological scale indicate the existence of so-called dark matter and dark energy \cite{Sahni2005}, which might mean that GR is invalid at this scale. For these reasons, since Einstein proposed the theory of GR, many experiments have tested the validity of this theory of gravity on various scales from sub-millimeter-scale tests in the laboratory to tests at solar system and cosmological scales \cite{Weinberg, Will2014,Will2014test,Hoyle2001,GPB2011,Sabulsky2019}. However, most of these experiments focus on the weak-field effects.

Gravitational wave (GW) is one of the most important predictions of GR, which can only be generated in strong gravitational fields and hardly interacts with matter, there is an excellent opportunity to test the theory of gravity in the strong-field regime. In recent years, with the detection of the first GW event from a binary black holes (BBH) system GW150914 \cite{GW150914} and other GW events \cite{GW170817,GW190521, GWTC1,GWTC2,GWTC3}, the theory and detection of GWs have attracted much attention. Numerous works have used GWs as a new probe for gravitational testing \cite{GW150914test,GW170817test,GWTC2test,GWTC3test}.

The BBH systems radiate GWs during the process of inspiraling to merging \cite{Will2014,MTW,Maggiore2008}. Because the binary has a very strong gravitational field in the pre-merger phase, the radiated GW may encode the distinction between the real gravitational theory and GR. Therefore, the detection of GWs generated from BBH is an essential aspect of the experiment of gravitational theory \cite{YunesppE2009,Li2012,Agathos2014}. With the improvement of sensitivities of GW detectors, future GW tests of gravitational theory impose higher requirements on the calculation of waveform templates \cite{Jaranowski2012}. Constructing such accurate templates has motivated significant research on post-Newtonian (PN) approximation method, which is applied to compute the GW waveforms in the regime where the two bodies have large separation. In the PN framework, the separation remains large with respect to the radii of both objects, thus the bodies can be regarded effectively as point particles.

The testing of GR by GW observations entails comparing the predictions of GW signals in GR and those in the alternative theories and constraining their differences by observations. Therefore, the choice of typical alternative gravitational theory \cite{BransDicke1961,ChernSimons2003,EDGB1996,Horndeski1974,Eling2004,MOG2015,XianGao2014,XianGao2019} and the calculation of GW waveforms in the theory have crucial roles. For example, the gravitational waveforms have been derived in Brans-Dicke (BD) gravity \cite{XingZhang2017BD,TanLiu2020BD,XiangZhao2019}, Einstein-aether theory \cite{ChaoZhang2020,KaiLin2019,XiangZhao2019}, screen modified gravity \cite{XingZhang2016ppNSMG,XingZhang2017SMG,TanLiu2018SMG,XingZhang2019SMG2}, and Horndeski theory \cite{Chowdhuri2022}. Current and future GW observations can constrain these gravitational theories with high accuracy \cite{XingZhangNSMGtest,RuiNiu2019,XingZhangSMGtest,RuiNiu2020,RuiNiu2021test,ChangfuShi2022,Carson2020,Chamberlain2017,Barausse2016,Perkins2021}.
% As people realized the inherent contradiction between GR and quantum theory \cite{Kiefer2007}, various modified gravitational theories have been proposed \cite{BransDicke1961,ChernSimons2003,EDGB1996,Horndeski1974,Eling2004,MOG2015,XianGao2014,XianGao2019,niupta}. 
Some kinds of modified theories, including parity-violating effects, have been studied in recent years \cite{Smith2008,Yunes2009CStest,Yunes2010PVtest,Mirshekari2012,WenZhao2020test,QiangWu2022,ChengGong2022,AnzhongWang2013,TaoZhu2013,JinQiao2020,TaoZhu2023,JinQiao2023,Boudet2023,Bombacigno2023}, like Chern-Simons (CS) gravity \cite{ChernSimons2003,Alexander2009} (consisting of dynamical and non-dynamical cases), ghost-free parity-violating theory \cite{JinQiao2019,WenZhao2020waveform,YifanWang2021,YifanWang2022,ZhichaoZhao2022}, Nieh-Yan modified teleparallel gravity \cite{MingzheLi2020NY,MingzheLi2021NY}, parity-violating symmetric teleparallel gravities \cite{MingzheLi2022STG} and the general spatial covariant gravities \cite{Takahashi2009,AnzhongWang2013,AnzhongWang2017,TaoZhu2013,XianGao2014,XianGao2019,TaoZhu2022,TaoZhu2023}. As a typical example, CS theory introduces a coupling term between the pseudo-scalar and Pontryagin density in the Einstein-Hilbert action, which causes the non-conservation of the CS topological current. One of the most important predictions of CS theory is that the amplitude of the left-hand circular polarization mode of GWs increases (or decreases) during the propagation while the amplitude for the right-hand mode decreases (or increases). This phenomenon is always called amplitude birefringence of GWs \cite{Alexander2009}.

% \footnote{In a more general ghost-free parity-violating theory, there are different propagation velocities for the left and right hand polarization modes, which is called velocity birefringence effect \cite{JinQiao2019,WenZhao2020waveform,wyf1,wyf2,other1}. Another theories, like Nieh-Yan modified teleparallel gravity \cite{MingzheLi2020NY,MingzheLi2021NY} and parity-violating symmetric teleparallel gravities \cite{MingzheLi2022STG} will only lead to velocity birefringence effect. This kind of parity-violating gravity has been tested through solar system experiments, double binary pulsar and GW detection \cite{Smith2008,Yunes2009CStest,Yunes2010PVtest,Mirshekari2012,WenZhao2020test,QiangWu2022,ChengGong2022,a1,a2,a3,a4}.}.

% Calculating the gravitational waveforms in modified gravities is necessary to constrain these modifications through GW detection. 
{\color{black}This work aims to derive the gravitational waveform in dynamical CS (dCS) gravity. The GW polarization modes are investigated in both generation and propagation processes.} Although the binary equation of motion (EOM) and PN waveform in the dCS theory are obtained by Refs.\,\cite{Yagi2012pn,Yagi2012gw,Yagi2016e}, the polarization of GWs is not studied, and the energy flux carried by tensor radiation is absent. The latter can result in a problematic parameterized post-Einsteinian (ppE) parameters.
% the effective density of spinning particle,
% the scalar fields of binary are treated as two magnetic dipoles, and then the interaction energy gives dCS modification to Newtonian gravity.
% Although this method is effective up to leading-order correction, more accurate calculation is necessary. Additionally.
In this work, we will recalculate the gravitational waveforms generated by the inspirals of BBH, which are different from Refs.\,\cite{Yagi2012pn,Yagi2012gw,Yagi2016e} in the following three aspects. First, we derive the EOM of BBH from the modified Mathisson-Papapetrou-Dixon (MPD) equation \cite{Bini2014}, a powerful tool to describe the motion of spinning point particle, which is based on the effective field theory (EFT) approach \cite{Steinhoff2011,Puetzfeld2015} and is extended by \cite{Loutrel2018,Loutrel2022} to study the motion and precession of spinning particles in dCS theory. Second, the investigation of GW polarization modes and improved energy flux are included. Third, the propagation effect, i.e., amplitude birefringence, is discussed. 
In conclusion, the improved ppE parameters and waveform including propagation effects are obtained in this work.
% Due to the parity-violating effect, dCS theory has no correction for spherically symmetric spacetime \cite{Grumiller2008}. That is, the Schwarzschild metric is still a solution to this theory. Previous works \cite{Konno2009,Yunes2009,Yagi2012} have given the slowly-rotating BH solution up to the spin-squared order. Therefore, only the GWs emitted by the binary system consisting of rotating BHs or neutron stars encode the violation from GR.

This paper is organized as follows. In Section \ref{sec:dCS}, we briefly review the dCS theory. In Section \ref{sec:EOM}, we focus on the spin-aligned case and obtain the EOM of BBHs up to 2PN approximation. Section \ref{sec:radiation} calculates the scalar and tensor radiation fields by multipole moment formulae. The polarization modes of generated GW are obtained through Newman-Penrose (NP) tetrad \cite{Will2014test,Wagle2019} in Section \ref{sec:polarization}. Furthermore, the energy flux carried by radiation and the orbital evolution are given in Section \ref{sec:flux}. In Section \ref{sec:FDwaveform}, we obtain the ready-used frequency-domain waveform. The birefringence effect of GWs when propagating in the cosmological background is discussed in Section \ref{sec:propagation}. Finally, we make a summary and discussion in Section \ref{sec:conslusion}.

Throughout the paper, we adopt the following conventions: We work in four dimensions with metric signature $(-,+,+,+)$, Latin indices $(a,b,c,\cdots,j,k, ...)$ in the index list represent spatial indices, whereas Greek indices $(\alpha,\beta,\cdots)$ represent spacetime indices, round brackets around indices represent symmetrization, square brackets represent anti symmetrization, $\partial_{\mu}$ represents a partial derivative, $\nabla_{\mu}$ represents a covariant derivative, $\Box^2_{\eta}\equiv\partial_{\mu}\partial^{\mu}$,  $\Box^2_{g}\equiv\nabla_{\mu}\nabla^{\mu}$, whereas $\bm{\nabla}^2\equiv\partial_{j}\partial_j$, the Einstein summation convention is employed, and we work in geometric units in which $c=G=1$, where $c$ is the speed of light in the vacuum and $G$ is the gravitational constant.

\section{Dynamical Chern-Simons Gravity}
\label{sec:dCS}
% In this section, we outline the basic knowledgement of dCS theory \cite{ChernSimons2003,Alexander2009,Yagi2012,Yagi2012pn,Loutrel2018}. Subsection \ref{subsec:action} reviews the action and the field equations of scalar and tensor fields. The slowly-rotating BH solution is introduced in subsection \ref{subsec:BH}. The perturbation equations are given in subsection \ref{subsec:perturbation} up to quadratic order of GR metric and linear order of deformation tensor and scalar perturbation. Finally, in preparation for deriving the EOM of binary BHs, the dCS-modified MPD equation is given in subsection \ref{subsec:MPD}.?????????????

\subsection{Action and Field Equations}
\label{subsec:action}
The full action of the dCS theory is \cite{Alexander2009}
\begin{equation}
\label{action}
\begin{aligned}
S=\int d^4x\sqrt{-g}\left[\frac{1}{16\pi}R
+\frac{\alpha}{4}\vartheta R\hat{R}
-\frac{\beta}{2}(\nabla_{\mu}\vartheta)(\nabla^{\mu}\vartheta)
+\mathcal{L}_{m}\right],
\end{aligned}
\end{equation}
where the gravity is described by a pseudo scalar field $\vartheta$ and the metric $g_{\mu\nu}$. As in Ref.\,\cite{Yagi2012pn}, we do not consider the potential of the scalar field. In Eq.\,(\ref{action}), $g$ is the determinant of the metric $g_{\mu\nu}$ and $R$ is the Ricci scalar. $\mathcal{L}_{m}$ is the Lagrangian density of the matter field. $\alpha$ and $\beta$ are the coupling parameters. $R\hat{R}\equiv(1/2)\varepsilon^{\rho\sigma\alpha\beta}R_{\nu\mu\rho\sigma}R^{\mu\nu}_{\ \ \alpha\beta}$ is the Potryagin density, with $R_{\nu\mu\rho\sigma}$ being the Riemann tensor and $\varepsilon^{\rho\sigma\alpha\beta}$ being the Levi-Civit\'{a} tensor defined in terms of the antisymmetric symbol $\epsilon^{\rho\sigma\alpha\beta}$ as $\varepsilon^{\rho\sigma\alpha\beta}=(1/\sqrt{-g})\epsilon^{\rho\sigma\alpha\beta}$.

The variation of the total action (\ref{action}) with respect to the metric $g_{\mu\nu}$ yields the modified field equations \cite{ChernSimons2003,Alexander2009},
\begin{equation}
\label{tensor-equation}
R_{\mu\nu}
-\frac{1}{2}g_{\mu\nu}R
+16\pi\alpha C_{\mu\nu}
=8\pi\left[T_{\mu\nu}^{(m)}+T_{\mu\nu}^{(\vartheta)}\right],
\end{equation}
where $R_{\mu\nu}$ is Ricci tensor and $C_{\mu\nu}$ is Cotton tensor defined as
\begin{equation}
\label{Ctensor}
C^{\mu\nu}=
-\varepsilon^{\rho(\mu|\alpha\beta}
\left[\nabla_{\alpha}R^{\nu)}_{\ \beta}\right]
(\nabla_{\rho}\vartheta)
-\hat{R}^{\kappa(\mu|\rho|\nu)}
(\nabla_{\kappa}\nabla_{\rho}\vartheta).
\end{equation}
Note that the Cotton tensor $C_{\mu\nu}$ is traceless, $g^{\mu\nu}C_{\mu\nu}=0$, and satisfies the Bianchi identity, $\nabla^{\mu}C_{\mu\nu}=0$. 
% {\color{red}{[C-tensor causes the amplitude birefringence during GW propagation, but has no contribution to GW generation]}}
$T_{\mu\nu}^{(m)}$ denotes the energy-momentum tensor (EMT) of the matter fields, and $T_{\mu\nu}^{(\vartheta)}$ represents the EMT of the coupled CS scalar field,
\begin{equation}
\label{EMTtheta}
T_{\mu\nu}^{(\vartheta)}=\beta
\left[(\nabla_{\mu}\vartheta)(\nabla_{\nu}\vartheta)
-\frac{1}{2}g_{\mu\nu}
(\nabla_{\alpha}\vartheta)(\nabla^{\alpha}\vartheta)\right].
\end{equation}

The equation of the scalar field can also be derived by variation the action (\ref{action}) with respect to the scalar field $\vartheta$, which leads to
\begin{equation}
\label{scalar-equation}
\beta\Box^2_{g}\vartheta
=-\frac{\alpha}{4}R\hat{R}.
\end{equation}

We would like to mention here that when the coupling constant $\beta=0$, the total action (\ref{action}) reduces to that of the non-dynamical CS gravity. In the non-dynamical case, the scalar field equation (\ref{scalar-equation}) becomes an additional differential constraint, i.e, the {\em Pontryagin constraint} on the space of allowed solutions,
\begin{equation}
R\hat{R}=0.
\end{equation}
In this paper we will not consider this case and only focus on the dCS gravity in which the parameter $\beta\neq 0$.

%CS theory is significantly different for $\beta=0$ and $\beta\neq0$ cases. The scalar field equation is meaningless for the former case, and the scalar field EMT vanishes. This theory is called non-dynamical CS theory. In this article, we focus on the dynamical case with $\beta\neq 0$, where scalar fields are distributed inhomogeneously in space and carry energy in the radiation process, thus affecting the evolution of the binary system.

\subsection{Slowly-Rotating Black Hole}
\label{subsec:BH}

One of important features of the dCS theory is the parity violation induced by the coupling between the scalar field $\vartheta$ and the Pontryagin density $R\hat{R}$. Thus, the Pontryagin density in general disappears in spherically symmetric spacetime. For this reason, the Schwarzschild black hole (BH) is still an exact solution to the dCS theory \cite{Grumiller2008}. The GWs radiated from binary Schwarzschild BHs are the same as those in GR \cite{Alexander2009}. Since the Pontryagin density normally is not vanishing for non-spherically symmetric gravitational systems, it is expected that the dCS theory can induce significant contributions to the GW waveform generated by binary spinning BHs. In this work, we focus on the gravitational radiation from binary spinning BHs. For this purpose, one has to first derive the spinning BH solution in dCS theory. Currently, the slowly rotating solution have been found within the small-coupling and small-spin approximation \cite{Yunes2009,Yagi2012}, which describes a slowly-rotating BH in the vacuum. The full metric $g_{\mu\nu}$ of the slowly-rotating BH solution and the scalar field profile $\vartheta(r)$ are expanded in terms of coupling constant $\alpha$ and dimensionless spin $\chi$ as follows,
\begin{equation}
\label{BH-metric-expand}
g_{\mu\nu}
=g_{\mu\nu}^{\rm(K)}
+\alpha^{\prime2}\chi' g^{(2,1)}_{\mu\nu}
+\alpha^{\prime2}\chi^{\prime2}g^{(2,2)}_{\mu\nu}
+\cdots
\end{equation}
\begin{equation}
\label{BH-scalar-expand}
\vartheta=\alpha'\chi'\vartheta^{(1,1)}
+\alpha'\chi^{\prime2}\vartheta^{(1,2)}
+\cdots.
\end{equation}
The bookkeeping parameters $\alpha'$ and $\chi'$ denote the expansion orders. In Boyer-Lindquist coordinate \cite{BoyerLindquist1967}, $(\tilde{t},\tilde{r},\tilde{\theta},\tilde{\varphi})$, $g^{(2,1)}_{\mu\nu}$, $g^{(2,2)}_{\mu\nu}$, $\vartheta^{(1,1)}$ and $\vartheta^{(1,2)}$ are undetermined functions of $\tilde{r}$ and $\tilde{\theta}$. The zero-order approximation $g_{\mu\nu}^{\rm(K)}$ is the Kerr metric \cite{Kerr1963}. The superscripts $(m,n)$ represent the expansion order of $\alpha'$ and $\chi'$. These undetermined functions can be solved by substituting these expansions, (\ref{BH-metric-expand}) and (\ref{BH-scalar-expand}), into the field equations, (\ref{tensor-equation}) and (\ref{scalar-equation}) \cite{Yagi2012}. Up to the required PN order, the scalar field profile at each order are given by
\begin{equation}
\label{BH-scalar-solution}
\vartheta^{(1,1)}
=-\frac{5}{8}\chi\frac{\alpha}{\beta}\frac{\cos\tilde{\theta}}{\tilde{r}^2},\quad\text{and}\quad
\vartheta^{(1,2)}=0. 
\end{equation}
And the metric field are
\begin{equation}
\label{BH-metric-solution-1}
g_{\tilde{t}\tilde{\varphi}}^{(2,1)}
=10\frac{\pi\alpha^2}{\beta m_0^4}\left(\frac{m_0}{\tilde{r}}\right)^4m_0\chi\sin^2\tilde{\theta},
\end{equation}
and
\begin{equation}
\label{BH-metric-solution-2}
g^{(2,2)}_{\tilde{t}\tilde{t}}=g^{(2,2)}_{\tilde{r}\tilde{r}}
=\frac{1}{\tilde{r}^2}g^{(2,2)}_{\tilde{\theta}\tilde{\theta}}
=\frac{1}{\tilde{r}^2\sin^2\tilde{\theta}}
g^{(2,2)}_{\tilde{\varphi}\tilde{\varphi}}
=\frac{201}{112}\frac{\pi\alpha^2}{\beta m_0^4}
\left(\frac{m_0}{\tilde{r}}\right)^3\chi^2(3\cos^2\tilde{\theta}-1).
\end{equation}
The scalar and metric solutions (\ref{BH-scalar-solution}),  (\ref{BH-metric-solution-1}), and (\ref{BH-metric-solution-2}) are of $\mathcal{O}(\chi)$ order and $\mathcal{O}(\chi^2)$ order, respectively. $m_0$ is the mass parameter of the slowly-rotating BH. Other components in the metric function are zero up to $\mathcal{O}(\alpha^2\chi^2)$ order.

%It should be noted that $\chi$ here is only a parameter (called bare spin) of the slowly-rotating BH solution, not the physical spin angular momentum. The relationship between bare and physical spin of dCS BH is \cite{Yagi2012}
%\begin{equation}
%\tilde{\chi}=\chi\left(1-\frac{709}{448}\frac{\pi\alpha^2}{\beta m_0^4}\right)
%\end{equation}
%In GR, where $\alpha=0$, the bare spin is the same as the physical spin. For observers, we can determine the physical spin of the BH by its effect on the surrounding spacetime. However, observers cannot tell whether this is the bare spin under GR or the dCS-corrected physical spin, because the bare spin is only an intrinsic parameter of the BH and cannot be measured.

\subsection{Scalar and Tensor Perturbation}
\label{subsec:perturbation}
The full metric is decomposed into a flat metric background $\eta_{\mu\nu}$ and a full metric perturbation $H_{\mu\nu}$, 
\begin{equation}
\label{metric-g}
g_{\mu\nu}=\eta_{\mu\nu}+H_{\mu\nu},
\end{equation}
where the perturbation $H_{\mu\nu}$ is further decomposed into the GR part $h_{\mu\nu}$ and a metric deformation away from GR $k_{\mu\nu}$ \cite{Yagi2012pn}, i.e.,
\begin{equation}
    H_{\mu\nu}=h_{\mu\nu}+k_{\mu\nu}.
\end{equation}
Accordingly, the inverse metric can be approximately written in the form of \cite{MTW}
\begin{equation}
\label{inverse-metric-g}
g^{\mu\nu}\approx\eta^{\mu\nu}-H^{\mu\nu}+H^{\mu\alpha}H^{\nu}_{\alpha},
\end{equation}
up to the second-order perturbation. Similar to the calculation of GW radiation from binary system in GR with the relaxed Einstein equation, it is convenient to define the standard trace-reversed metric perturbation $\bar{H}^{\mu\nu}$ as \cite{Landau,Maggiore2008},
\begin{equation}
\label{metric-H}
\bar{H}^{\mu\nu}\equiv
\eta^{\mu\nu}-\sqrt{-g}g^{\mu\nu},
\end{equation}
and here one requires $\bar{H}^{\mu\nu}$ satisfying the Lorenz gauge,
\begin{equation}
\label{Lorenz}
\partial_{\mu}\bar{H}^{\mu\nu}=0.
\end{equation}
At the same time, the trace-reversed metrics $\bar{h}_{\mu\nu}$ and $\bar{k}_{\mu\nu}$ are defined as
\begin{equation}
\bar{h}^{\mu\nu}=h^{\mu\nu}
-\frac{1}{2}\eta^{\mu\nu}h
+\frac{1}{2}hh^{\mu\nu}
+\frac{1}{4}\eta^{\mu\nu}h_{\alpha\beta}h^{\alpha\beta}
-\frac{1}{8}\eta^{\mu\nu}h^2
-h^{\mu\alpha}h_{\alpha}^{\nu},
\end{equation}
and
\begin{equation}
\bar{k}^{\mu\nu}=k^{\mu\nu}-\frac{1}{2}\eta^{\mu\nu}k,
\end{equation}
respectively. The traces of GR perturbation then given by $h\equiv\eta^{\mu\nu}h_{\mu\nu}$ and $\bar{h}\equiv\eta^{\mu\nu}\bar{h}_{\mu\nu}$, and the traces of metric deformation are defined as $k\equiv\eta^{\mu\nu}k_{\mu\nu}$ and $\bar{k}\equiv\eta^{\mu\nu}\bar{k}_{\mu\nu}$.

As shown in Section \ref{subsec:BH}, the scalar and gravitational fields should be of order $\mathcal{O}(\alpha)$ and $\mathcal{O}(\alpha^2)$, respectively. Therefore, to derive the scalar equation, the terms who is of order $\mathcal{O}(\vartheta h)$ and $\mathcal{O}(h^2)$ are retained, and that of order $\mathcal{O}(hk)$ are removed. And to derive the tensor equation, the perturbation order is of $\mathcal{O}(\vartheta^2)$ and $\mathcal{O}(k)$. 
% Additionally, we need to take the second-order GR perturbation, $\mathcal{O}(h^2)$, into consideration to obtain the 2PN near-zone solution and waveforms.

With the above setups, one is able to get the second-order perturbated equations by substituting decomposition (\ref{metric-g}) into Eqs.\,(\ref{tensor-equation}) and (\ref{scalar-equation}). For the scalar field equation, it becomes \cite{Yagi2016e}
\begin{equation}
\label{scalar-perturbation}
\Box^2_{\eta}\vartheta=-16\pi\sigma,
\end{equation}
where the source term $\sigma$ is given by
\begin{equation}
\label{scalar-source}
16\pi\sigma
=\frac{1}{2}(\partial_{\mu}h)(\partial^{\mu}\vartheta)
-(\partial_{\mu}h^{\mu\nu})(\partial_{\nu}\vartheta)
-h^{\mu\nu}(\partial_{\mu}\partial_{\nu}\vartheta)
+\frac{\alpha}{4\beta}
\epsilon^{\rho\sigma\alpha\beta}
(\partial_{\rho}\partial^{\lambda}h_{\mu\sigma})
(\partial_{\alpha}\partial^{\mu}h_{\lambda\beta}
-\partial_{\alpha}\partial_{\lambda}h^{\mu}_{\beta}).
\end{equation}

As we have mentined in the above, the full metric perturbation $H_{\mu\nu}$ is divided into two parts: the GR part $h_{\mu\nu}$ and metric deformation $k_{\mu\nu}$ away from GR. The former satisfies the quadratic-linearized relaxed Einstein equation \cite{Landau,Maggiore2008}, while the latter one satisfies \cite{Yagi2016e}
\begin{equation} 
\label{tensor-perturbation}
\Box^2_{\eta}\bar{k}_{\mu\nu}
=-16\pi K_{\mu\nu},
\end{equation}
where the source term $K_{\mu\nu}$ is 
\begin{equation}
\label{tensor-source}
K_{\mu\nu}
=kT_{\mu\nu}^{(m)}
+\left(1+h\right)\delta T_{\mu\nu}^{(m)}
+T_{\mu\nu}^{(\vartheta)}
+\frac{1}{16\pi}\tilde{\Lambda}_{\mu\nu}
-2\alpha\tilde{C}_{\mu\nu}.
\end{equation}
Here the quantity $\delta T_{\mu\nu}^{(m)}$ stands for the perturbation to the EMT of the matter fields, and the tensor $\tilde{\Lambda}^{(2)}_{\mu\nu}$ is defined as
\begin{equation}
\label{Lambda-source}
\tilde{\Lambda}_{\mu\nu}
\approx16\pi(-g)\tilde{t}_{\mu\nu}^{\rm LL}
-(\partial^{\alpha}\partial^{\beta}\bar{h}_{\mu\nu})\bar{k}_{\alpha\beta}
-(\partial^{\alpha}\partial^{\beta}\bar{k}_{\mu\nu})\bar{h}_{\alpha\beta}
+(\partial^{\beta}\bar{h}_{\alpha\mu})(\partial^{\alpha}\bar{k}_{\beta\nu})
+(\partial^{\beta}\bar{k}_{\alpha\mu})(\partial^{\alpha}\bar{h}_{\beta\nu}),
\end{equation}
up to the linear order of metric deformation. The quantity $\tilde{t}^{\rm LL}_{\mu\nu}$ in the expression of $\tilde{\Lambda}^{(2)}_{\mu\nu}$ denotes the linearized Landau-Lifshitz energy-momentum pseudotensor, which is given by \cite{Landau}
\begin{equation}
\label{LLEMT-source}
\begin{aligned}
16\pi(-g)\tilde{t}^{\rm LL}_{\mu\nu}
&=(\partial_{\alpha}\bar{h}_{\mu}^{\lambda})(\partial^{\alpha}\bar{k}_{\lambda\nu})
+(\partial_{\alpha}\bar{k}_{\mu}^{\lambda})(\partial^{\alpha}\bar{h}_{\lambda\nu})
+\eta_{\mu\nu}(\partial_{\beta}\bar{h}^{\alpha\lambda})(\partial_{\alpha}\bar{k}_{\lambda}^{\beta})\\
&-(\partial_{\mu}\bar{h}^{\beta\lambda})(\partial_{\beta}\bar{k}_{\lambda\nu})
-(\partial_{\mu}\bar{k}^{\beta\lambda})(\partial_{\beta}\bar{h}_{\lambda\nu})
-(\partial_{\nu}\bar{h}^{\beta\lambda})(\partial_{\beta}\bar{k}_{\lambda\mu})
-(\partial_{\nu}\bar{k}^{\beta\lambda})(\partial_{\beta}\bar{h}_{\lambda\mu})\\
&+(1/2)(\partial_{\mu}\bar{h}^{\lambda\sigma})(\partial_{\nu}\bar{k}_{\lambda\sigma})
+(1/2)(\partial_{\mu}\bar{k}^{\lambda\sigma})(\partial_{\nu}\bar{h}_{\lambda\sigma})
-(1/4)(\partial_{\mu}\bar{h})(\partial_{\nu}\bar{k})\\
&-(1/4)(\partial_{\mu}\bar{k})(\partial_{\nu}\bar{h})
-(1/2)\eta_{\mu\nu}(\partial_{\alpha}\bar{h}^{\lambda\sigma})(\partial^{\alpha}\bar{k}_{\lambda\sigma})
+(1/4)\eta_{\mu\nu}(\partial_{\alpha}\bar{h})(\partial^{\alpha}\bar{k}).
\end{aligned}
\end{equation}
Up to the linear order of the scalar field $\vartheta$ and the GR perturbation $h_{\mu\nu}$, the C-tensor becomes
\begin{equation}
\label{C-tensor-linear}
\tilde{C}_{\mu\nu}
=\frac{1}{2}
\eta_{(\mu|\lambda}\epsilon^{\rho\lambda\alpha\beta}
\partial_{\alpha}(\Box^2_{\eta}h_{\nu)\beta})
(\partial_{\rho}\vartheta)
-\frac{1}{2}
\eta_{(\nu|\lambda}\epsilon^{\rho\lambda\alpha\beta}
\left(\partial_{\alpha}\partial_{\mu)}h^{\kappa}_{\beta}
-\partial_{\alpha}\partial^{\kappa}h_{\beta\mu}\right)
(\partial_{\kappa}\partial_{\rho}\vartheta).
\end{equation}

The source $\delta T_{\mu\nu}^{(m)}$ in Eq.\,(\ref{tensor-source}) contains contributions proportional to the density, pressure and velocity of matter, which has \emph{compact support}, that is, is nonzero only in finite spatial regions where the matter resides. Inversely, other sources in Eq.\,(\ref{tensor-source}) contain contributions proportional to products of fields and their derivatives. These contributions extend over all spacetime, although they generally decrease with distance from the sources \cite{Will2014test}. Differently, the source of scalar field (\ref{scalar-source}) only involves non-compact support terms, while the compact-support terms will be derived from the EFT approach (see subsection \ref{subsec:MPD}). The near-zone solution and the motion of the binary system are mainly determined by the compact support terms, while the gravitational radiation is sourced from both the compact and non-compact support terms.

\subsection{Modified MPD Equations}
\label{subsec:MPD}
In this work, the BHs are modelled as a spinning point mass using the EFT method extended by Ref.\,\cite{Loutrel2018} to dCS gravity. The motion of BBH is influenced by the mass and the spin angular momentum (SAM) of the point particles. By considering a set of symmetries associated with theory, specifically (1) re-parameterization invariance of proper time $\tau$, (2) Lorentz invariance, and (3) diffeomorphism invariance, one can write the quadrupole-approximated Lagrangian of a spinning point mass as \cite{Steinhoff2011,Loutrel2018}
\begin{equation}
\label{Lagrangian}
L_{m}(u^{\mu},\Omega^{\mu\nu},R_{\alpha\beta\gamma\delta})
=p_{\mu}u^{\mu}
+\frac{1}{2}S_{\alpha\beta}\Omega^{\alpha\beta}
-\frac{1}{6}J^{\alpha\beta\gamma\delta}R_{\alpha\beta\gamma\delta}.
\end{equation}
In Eq.\,(\ref{Lagrangian}), $u_{\mu}$, $p_{\mu}$, $\Omega^{\alpha\beta}$ and $S^{\alpha\beta}$ are the 4-velocity, 4-momentum, angular velocity tensor and SAM tensor, respectively. The residual degree of freedom (DOF) of the SAM tensor is removed by the Tulczyjew spin supplementary condition \cite{Blanchet2014,Loutrel2018}, $S^{\mu\nu}p_{\mu}=0$. The Dixon's quadrupole moment is defined as \cite{Steinhoff2011,Bini2014,Puetzfeld2015,Loutrel2018}
\begin{equation}
\label{Dixon}
J^{\alpha\beta\gamma\delta}\equiv-\frac{3}{m_0}
u^{[\alpha}S^{\beta]\lambda}S_{\ \ \lambda}^{[\gamma}u^{\delta]}
=\frac{3}{m_0}\left(u^{[\alpha}S^{\beta]}S^{[\gamma}u^{\delta]}
-S^2u^{[\alpha}g^{\beta][\gamma}u^{\delta]}\right),
\end{equation}
where $S^2\equiv S^{\mu}S_{\mu}$ and $S^{\mu}$ is the spin 4-vector defined as $S^{\mu}\equiv-(1/2)\epsilon^{\alpha\beta\gamma\mu}u_{\gamma}S_{\alpha\beta}$, automatically satisfying the Tulczyjew condition. Then, the residual DOF is removed by gauge condition, $S^{\mu}u_{\mu}=0$.

However, dCS theory brings additional modification to the Lagrangian \cite{Loutrel2018}. The scalar field modifies the Lagrangian (\ref{Lagrangian}), and the slowly-rotating BH metric described by Eqs.\,(\ref{BH-scalar-solution}-\ref{BH-metric-solution-2}) violates the standard Kerr metric, an additional contribution to the Newtonian gravitational potential. By requiring the Lagrangian to be invariant under (1) a shift of the scalar field and (2) a parity transformation, the modified Lagrangian function is finally given by
\begin{equation}
\label{Lagrangian-modification}
L_{m}(u^{\mu},\Omega^{\mu\nu},
R_{\alpha\beta\gamma\delta},\nabla^{\alpha}\vartheta)
=p_{\mu}u^{\mu}
+\frac{1}{2}S_{\alpha\beta}\Omega^{\alpha\beta}\\
-\frac{1}{6}(1+\delta C_{Q})
J^{\alpha\beta\gamma\delta}R_{\alpha\beta\gamma\delta}
+\frac{1}{m_0^2}\delta C_{\vartheta}\hat{S}_{\mu}^{\ \alpha}
u^{\mu}(\nabla_{\alpha}\vartheta),
\end{equation}
where the correction constants, $\delta C_{\vartheta}$ and $\delta C_{Q}$, are determined by comparing the near-zone and the slowly-rotating BH solution. The dual tensor of SAM is $\hat{S}_{\mu\nu}\equiv(1/2)\epsilon_{\mu\nu}^{\ \ \lambda\rho}S_{\lambda\rho}$.

From the modified Lagrangian (\ref{Lagrangian-modification}), the modified MPD equations are given by \cite{Loutrel2018}
\begin{subequations}
\label{MPD}
\begin{align}
\label{MPD-EOM}
u^{\alpha}\nabla_{\alpha}p_{\mu}
&=\frac{1}{2}S_{\alpha\beta}
u^{\lambda}R^{\alpha\beta}_{\ \ \lambda\mu}
+\frac{1}{m_0^2}\delta C_{\vartheta}\hat{S}_{\lambda}^{\ \alpha}
u^{\lambda}(\nabla_{\mu}\nabla_{\alpha}\vartheta)
-\frac{1}{6}(1+\delta C_{Q})J^{\alpha\beta\gamma\delta}
\nabla_{\mu}R_{\alpha\beta\gamma\delta},\\
\label{MPD-EOP}
u^{\alpha}\nabla_{\alpha}S_{\mu\nu}
&=2p_{[\mu}u_{\nu]}+\frac{2}{m_0^2}\delta C_{\vartheta}
u^{\lambda}\hat{S}_{\lambda[\mu}(\nabla_{\nu]}\vartheta)
+\frac{4}{3}(1+\delta C_{Q})
J^{\rho\lambda\alpha}_{\ \ \ \ [\mu}
R_{\nu]\alpha\rho\lambda}.
\end{align}
\end{subequations}
These two equations reduce to standard MPD equations by setting $\delta C_{\vartheta}$ and $\delta C_{Q}$ to zeros \cite{Bini2014}. In a BBH system, Eq.\,(\ref{MPD-EOM}) determines the acceleration of the particle in dCS gravity and Eq.\,(\ref{MPD-EOP}) is the spin-precession equation for the particle.

From the modified Lagrangian (\ref{Lagrangian-modification}), the modified EMT of a spinning point mass is written as \cite{Blanchet2014,Loutrel2018}
\begin{equation}
\label{spinning-EMT}
T_{(m)}^{\mu\nu}=\int \frac{d\tau}{\sqrt{-g}}\left[m_0u^{\mu}u^{\nu}\delta^{(4)}
-\nabla_{\lambda}(S^{\lambda(\mu}u^{\nu)}\delta^{(4)})\\
-\frac{2}{3}(1+\delta C_{Q})
\nabla_{\alpha}\nabla_{\beta}(J^{\mu\alpha\beta\nu}\delta^{(4)})\right]
(x^{\mu}-z^{\mu}(\tau)).
\end{equation}
The three terms in the above EMT tensor act as mass, spin and quadrupole source terms, respectively. $x^{\mu}$ is the coordinate of a point and $z^{\mu}(\tau)$ is the trajectory of the point mass. Some terms describing scalar-spin coupling and quadrupole-spacetime coupling are omitted because they do not contribute to the gravitational field after regularization \cite{Loutrel2018}.

EFT shows that the point-particle approximation brings an additional compact support term to the scalar field equation (\ref{scalar-equation}) \cite{Yagi2016e,Loutrel2018}, 
\begin{equation}
\label{scalar-compact-source}
\rho_{\vartheta}=-\frac{\delta C_{\vartheta}}{4\pi\beta m_0^2}
\int d\tau\hat{S}_{\mu}^{\ \alpha}u^{\mu}
\nabla_{\alpha}
\left[\frac{\delta^{(4)}(x^{\mu}-z^{\mu})}{\sqrt{-g}}\right].
\end{equation}
This quantity acts as an effective dipolar source for the scalar field. $\delta^{(4)}(x^{\mu}-z^{\mu})$ is four-dimensional Dirac delta function. Then the field equation (\ref{scalar-perturbation}) is rewritten as
\begin{equation}
\label{scalar-perturbation-modification}
\Box^2_{\eta}\vartheta
=-16\pi\sigma+4\pi\rho_{\vartheta}.
\end{equation}
This compact support term will play an important role in the near-zone solution and radiated waveform.

\section{Equation of Motion}
\label{sec:EOM}
% We derive the EOM of binary BH system for spin-aligned circular orbits up to 2PN approximation. At this time, EOP becomes uα∇αSμν = 0, indicating that the SAM of BHs is conserved. For preparation, the scalar solution and PN metric in the near-zone (defined in subsection III A) are given in subsections III B and III C. And we will present the 2PN EOM in subsection III D. The modified relative EOM, modified Keplerian law and modified conserved energy are give in subsections III E, III F and III G, respectively.

In this Section, The EOM of the BBH system with quasi-circular orbit is derived up to 2PN approximation. The SAMs of BHs are assumed to be aligned to the orbital angular momentum of the system. At this time, the precession equation becomes $u^{\alpha}\nabla_{\alpha}S_{\mu\nu}=0$, indicating that the SAM of BHs is conserved. First, the approximate solution of the scalar field and the PN metric in the near zone are derived for preparation. Second, the 2PN EOM of BBH is presented. Third, the modified relative EOM, Kepler's third law, and conserved energy are given in subsequent subsections.

\subsection{Zones}
\label{subsec:zones}
This paper will discuss the generation and propagation of GWs on multiple scales. For the convenience of later discussions, we divide the whole space into four zones: \emph{inner zone}, \emph{near zone}, {\color{black}\emph{wave zone}, and \emph{propagation zone}}. The inner zones are centered on each object of the binary system. Their radiuses are much smaller than the typical orbital radius of the binary system. The gravitational fields in these zones are too strong to apply the PN approximation and linear GW theory. The near zone is centered on the binary's center of mass, whose radius is comparable with the typical wavelength of emitted GWs.
%The gravitational field in this zone is the linear combination of that of each object. The background metric determines the binary motion and radiation source.
The wave zone is also centered on the binary, whose radius is about several times the wavelength but far less than the distance between galaxies. The background can be considered flat, and GW propagates freely in this zone. The propagation zone includes all other regions far away from the source. The cosmological background must be considered when GWs propagate in this zone. In dCS theory, the homogeneous background scalar field induces the amplitude birefringence effect of GWs.

\subsection{Near-Zone Solution: Scalar}
\label{subsec:NZscalar}
The scalar field generated by a slowly-rotating BH acts like a magnetic dipole. For simplification, we define the scalar dipole moment (SDM) of one BH in a binary system (the $A$-th BH) as
\begin{equation}
\label{SDM}
\mu^{\alpha}_A\equiv\frac{\delta C_{\vartheta}}{4\pi\beta m_A^2}S_A^{\alpha},
\end{equation}
where $m_{A}$ and $S_{A}^{\alpha}$ are the mass and SAM of the $A$-th BH. The SDM vector is constrained by gauge condition $\mu_A^{\alpha}u^A_{\alpha}=0$. In the Newtonian limit, the proper time of a particle $\tau$ is just the coordinate time $t$. Thus the time component of 4-velocity is $u^A_{0}=-1$, and the spatial components are replaced by $v_A^{i}\equiv u_A^{i}/u_A^{0}$. Such that the time component of SDM is $\mu_A^{0}=\mu_A^{i}v^A_{i}$. Then the compact-support source $\rho_{\vartheta}$ (\ref{theta-rho-def}) is written as \cite{Yagi2016e}
\begin{equation}
\label{scalar-compact-source-3}
\rho_{\vartheta}=\rho_{1}+\rho_{2}+\rho_{3}.
\end{equation}
The three terms in Eq.\,(\ref{scalar-compact-source-3}) are
\begin{subequations}
\label{scalar-compact-source-123}
\begin{align}
\label{scalar-compact-source-part1}
\rho_{1}&=\sum_{A}\mu_A^{i}
\partial_{i}\delta^{(3)}(\mathbf{x}-\mathbf{z}_A(t))
\sim\mathcal{O}(1/c^4),\\
\label{scalar-compact-source-part2}
\rho_{2}&=\sum_A\mu_A^{i}\dot{v}_A^{i}\delta^{(3)}(\mathbf{x}-\mathbf{z}_A(t))
\sim\mathcal{O}(1/c^6),\\
\label{scalar-compact-source-part3}
\rho_{3}&=-\sum_{A}
\mu_A^{i}v_A^{i}v_A^{j}\partial_{j}
\delta^{(3)}(\mathbf{x}-\mathbf{z}_A(t))
\sim\mathcal{O}(1/c^6),
\end{align}
\end{subequations}
where $\mathbf{x}$ and $\mathbf{z}_A$ are the spatial components of 4-vectors $x^{\mu}$ and $z_A^{\mu}$. In above equations, $c$ is the light speed in vacuum, indicating the magnitude of these terms. $\delta^{(3)}(\mathbf{x}-\mathbf{z})$ is three-dimensional Dirac delta function. With the help of the decomposition (\ref{scalar-compact-source-123}), the near-zone scalar solution of BBH is mainly sourced by the first term $\rho_{1}$ (\ref{scalar-compact-source-part1}). We finally obtain
\begin{equation}
\label{scalar-near-zone}
\vartheta^{\rm(B)}
\approx\int\rho_{1}(t,\mathbf{x}')
\frac{1}{|\mathbf{x}-\mathbf{x}'|}d^3\mathbf{x}'
=\sum_A\frac{\hat{\bm{n}}_A\cdot\bm{\mu}_A}{r_A^2},
\end{equation}
where the distance between point $\mathbf{x}$ and $A$-th BH is defined as $r_A\equiv|\mathbf{x}-\mathbf{z}_A|$, and the unit directional vector is defined as $\hat{\bm{n}}_A\equiv(\mathbf{x}-\mathbf{z}_A)/r_{A}$. 

On the other hand, the scalar field for the binary system is now just the sum of the individual contributions, specifically
\begin{equation}
\label{scalar-near-zone-summation}
\vartheta^{\rm (B)}
=-\frac{5}{8}\sum_{A}\frac{\alpha}{\beta m_A^2}
\frac{\hat{\bm{n}}_{A}\cdot\hat{\bm{S}}_{A}}{r_{A}^2}.
\end{equation}
Comparing Eqs.\,(\ref{scalar-near-zone}) and (\ref{scalar-near-zone-summation}), we can determine the value of the correction constant,
\begin{equation}
\label{delta-C-vartheta}
\delta C_{\vartheta}=-\frac{5}{2}\pi\alpha.
\end{equation}
The spatial components of SDM is re-expressed as
\begin{equation}
\label{ESC-2}
\bm{\mu}_{A}
=-\frac{5}{8}\frac{\alpha}{\beta m_A^2}\bm{S}_{A}.
\end{equation}
The SDM of a slowly-rotating BH is similar to the magnetic dipole moment produced by electron spin \cite{Yagi2012pn}, providing the spin-spin coupling between two BHs. This completes the computation of the near-zone scalar field of the binary system.

\subsection{Near-Zone Solution: Tensor}
\label{subsec:NZtensor}
We turn to investigate the near-zone solution of the metric. Up to 2PN order, the PN metric is given by \cite{Blanchet1995,Blanchet1998,Blanchet2001,Blanchet2002,Blanchet2014}
\begin{subequations}
\label{hh-near-zone}
\begin{align}
\label{h00-near-zone}
h_{00}^{\rm(B)}&=2\bar{U}-2\bar{U}^2
+8\left[\bar{X}-\bar{V}_i\bar{V}_i+(1/6)\bar{U}^3\right],\\
\label{h0i-near-zone}
h_{0i}^{\rm(B)}&=-4\bar{V}_{i}-8\bar{R}_{i},\\
\label{hij-near-zone}
h_{ij}^{\rm(B)}&=2\delta_{ij}\bar{U}
+\left(2\delta_{ij}\bar{U}^2+4\bar{W}_{ij}\right).
\end{align}
\end{subequations}
In Eq.\,(\ref{hh-near-zone}), several potentials $\bar{\mathbf{X}}=\{\bar{U}, \bar{V}_i, \bar{X}, \bar{R}_i, \bar{W}_{ij}\}$ are defined as the integrations of source terms (\ref{tensor-source}). The potential $\bar{U}$ reduces to the Newtonian potential at leading order. Because the matter EMT (\ref{spinning-EMT}) contains mass, spin, and quadrupole parts, these potentials are also correspondingly divided into these three parts, i.e. $\bar{\mathbf{X}}=\bar{\mathbf{X}}^{\rm(M)}+\bar{\mathbf{X}}^{\rm(S)}+\bar{\mathbf{X}}^{\rm(Q)}$, which are not shown here. The readers can find the explicit expressions of $\bar{\mathbf{X}}^{\rm(M)}$ in Ref.\,\cite{Blanchet1998}, $\bar{\mathbf{X}}^{\rm(S)}$ in Ref.\,\cite{Faye2006}, and $\bar{\mathbf{X}}^{\rm(Q)}$ in Refs.\,\cite{Buonanno2013,Bohe2015}. For instance, the monopole, dipole, and quadrupole sectors of Newtonian potential are 
\begin{equation}
\bar{U}^{\rm(M)}
=\sum_{A}\frac{m_{A}}{r_{A}}
+\text{(higher PN order without spin effects)}
\sim\mathcal{O}(1/c^2),
\end{equation}
\begin{equation}
\bar{U}^{\rm(S)}
=-2\varepsilon_{ijk}\sum_{A}v_{A}^{i}S_{A}^{j} \partial_{k}\left(\frac{1}{r_{A}}\right)
\sim\mathcal{O}(1/c^5),
\end{equation}
and
\begin{equation}
\bar{U}^{\rm(Q)}
=-\frac{3}{2}\sum_{A}
\left(\frac{m_A}{r_{A}}\right)^3\frac{1}{m_{A}^4}
\left[(\hat{\bm{n}}_{A}\cdot\bm{S}_{A})^2-\frac{1}{3}(\bm{S}_{A}\cdot\bm{S}_{A})\right]
\sim\mathcal{O}(1/c^6).
\end{equation}
As we have shown, the quadratic-spin sector of 2PN Newtonian potential in GR is proportional to $S_{A}^2/r_{A}^3$.

The lowest-order dCS modification to the metric perturbation (\ref{h00-near-zone}) is just
\begin{equation}
\label{U-plus-delta-U}
\bar{U}\rightarrow\bar{U}+\delta U.
\end{equation}
$\delta U$ is the retarded potential integration of the non-GR sector of matter EMT in Eq.\,(\ref{spinning-EMT}). The $00$ component of the leading-order contribution is
\begin{equation}
\label{delta-T00}
\delta T_{00}^{(m)}=\frac{2}{3}
\sum_A\delta C_{Q}J_A^{i0k0}
\partial_{i}\partial_{k}\delta^{(3)}(\mathbf{x}-\mathbf{z}_A(t)).
\end{equation}
Thus, by integrating the EMT (\ref{delta-T00}), we eventually find
\begin{equation}
\label{delta-U}
\begin{aligned}
\delta U\equiv\int\frac{\delta T^{(m)}_{00}}{|\mathbf{x}-\mathbf{x}'|}d^3\mathbf{x}'
&=-\frac{3}{2}\sum_A\frac{\delta C_Q}{m_A}
\left(S_A^{i}S_A^{j}-\frac{1}{3}\delta_{ij}S_A^{k}S_A^{k}\right)
\frac{\hat{n}_A^i\hat{n}_A^j}{r_A^3}\\
&=-\frac{3}{2}\sum_A\frac{\delta C_{Q}}{m_A^4}
\left(\frac{m_{A}}{r_{A}}\right)^3
\left[(\hat{\bm{n}}_{A}\cdot\bm{S}_{A})^2-\frac{1}{3}(\bm{S}_{A}\cdot\bm{S}_{A})\right]
\sim\mathcal{O}(1/c^6).
\end{aligned}
\end{equation}
The modified potential function $\delta U$ in Eq.\,(\ref{delta-U}) is similar to quadrupole gravitational potential $\bar{U}^{\rm(Q)}$, which contains $\mathcal{O}(S^2)$ terms and enters 2PN correction. The near-zone metric deformation tensors are 
\begin{equation}
k^{\rm(B)}_{00}=2\delta U,\quad\text{and}\quad k^{\rm(B)}_{0i}=k^{\rm(B)}_{ij}=0.
\end{equation}
at the leading PN order. We find that the $\delta U$ is proportional to $\delta C_{Q}\cdot S_{A}^2/r_{A}^3$, which is similar to $\bar{U}^{\rm(Q)}$ except a constant $\delta C_{Q}$, which is different for each object. Such that, the quadratic-spin effect and the dCS modification are both 2PN higher than the Newtonian potential, $m_{A}/r_{A}$. Other potential in the PN metric, $\{\bar{V}_{i}, \bar{X}, \bar{R}_{i}, \bar{W}_{ij}\}$, are higher PN-order effects relative to the Newtonian potential. The dCS modification to these potential functions, $\{\delta{V}_{i}, \delta{X}, \delta{R}_{i}, \delta{W}_{ij}\}$, are 2PN effects relative to leading terms of $\{\bar{V}_{i}, \bar{X}, \bar{R}_{i}, \bar{W}_{ij}\}$. Therefore, these corrections enter higher order correction. Accordingly, only the correction to the potential $\bar{U}$ is necessary to be considered when deriving the near-zone solution and EOM.

At the same time, the modified potential $\delta U$ also can be defined from the slowly-rotating BH solution as $g_{\tilde{t}\tilde{t}}=-1+2(\bar{U}+\delta U)$, with $g_{\tilde{t}\tilde{t}}$ being shown in Eq.\,(\ref{BH-metric-solution-2}). In the harmonic coordinates, which are identical to the Boyer-Lindquist coordinates at leading PN order \cite{Loutrel2018}, the potential is written as
\begin{equation}
\label{delta-U-BH}
\delta U
=-\frac{201}{224}\sum_{A}
\frac{\pi\alpha^2}{\beta m_{A}^4}
\chi_A^2\frac{m_A^3}{\tilde{r}_A^3}
\left(3\cos^2\tilde{\theta}_A-1\right)
=3\sum_{A}\delta Q_A^{ij}
\frac{\hat{n}_A^{i}\hat{n}_A^{j}}{\tilde{r}_A^{3}},
\end{equation}
where $\delta Q_A^{ij}$ is the quadrupole moment tensor of the $A$-th BH induced by dCS modification, defined by
\begin{equation}
\label{delta-Q}
\delta Q_A^{ij}
=\frac{201}{224}\frac{\pi\alpha^2}{\beta m_{A}^4}
m_{A}^3\chi_{A}^2
\left(\begin{array}{ccc}
-1/3 & 0 & 0\\
0 & -1/3 & 0\\
0 & 0 & 2/3
\end{array}\right).
\end{equation}
To find the value of constant $\delta C_{Q}$, we match the modified potential (\ref{delta-U-BH}) and Eq.\,(\ref{delta-U}) in the binary reference system, taking the center of mass of the binary system as the origin and the vertical direction of the orbital plane as the $z$-axis. Transforming the quadrupole moment tensors $\delta Q_A^{ij}$ (\ref{delta-Q}) from the Boyer-Lindquist coordinates to the binary reference system, the modified potential is now written as
\begin{equation}
\label{delta-U-BH-binary-reference}
\delta U
=\frac{603}{3584}\sum_A\frac{\zeta_{A}}{m_A^4}
\left(\frac{m_A}{r_A}\right)^3
\left[(\hat{\bm{n}}_{A}\cdot\bm{S}_{A})^2-\frac{1}{3}(\bm{S}_{A}\cdot\bm{S}_{A})\right].
\end{equation}
The redefined dimensionless coupling constant is defined as $\zeta_A=16\pi\alpha^2/\beta m_A^4$. Comparing Eq.\,(\ref{delta-U}) with (\ref{delta-U-BH-binary-reference}), we find that the exact result of modification constant $\delta C_{Q}$ is
\begin{equation}
\label{delta-C-Q}
\delta C_{Q}=-\frac{201}{1792}\zeta_{A},
\end{equation}
which depends on the typical mass of each object.

So far, the near-zone metric and scalar field are completely determined. On the one hand, the near-zone fields are the sum of the contributions from the individual objects. On the other hand, the near-zone solutions provide a way to find the undetermined constants in EFT and modified MPD equations.

\subsection{2PN EOM}
\label{subsec:2PN}
The modified MPD equations (\ref{MPD}) describe the motion and spin precession of BBH in dCS gravity. As we have mentioned, the SAM of each object is conserved for spin-aligned orbits. Thus, we only focus on the first equation (\ref{MPD-EOM}), which describes how the SAMs, scalar fields, and Dixon's quadrupole moments modify the accelerations. Four kinds of forces appear in the EOM, connection ($\Gamma$) force, spin-orbit (SO) force, dipole-dipole (DD) force, and monopole-quadrupole (MQ) force, defined as
\begin{subequations}
\label{force}
\begin{align}
\label{Gamma-force}
Y_{(\Gamma)}^{\mu}&\equiv
-m\Gamma_{\lambda\rho}^{\mu}u^{\lambda}u^{\rho},\\
\label{SO-force}
Y_{\rm(SO)}^{\mu}&\equiv-\frac{1}{2}\epsilon_{\alpha\beta\gamma\delta}
u^{\gamma}u^{\lambda}S^{\delta}R^{\alpha\beta\mu}_{\ \ \ \ \lambda},\\
\label{DD-force}
Y_{\rm(DD)}^{\mu}&\equiv\frac{1}{m^2}\delta C_{\vartheta}S^{\alpha}
(\nabla^{\mu}\nabla_{\alpha}\vartheta),\\
\label{MQ-force}
Y_{\rm(MQ)}^{\mu}&\equiv-\frac{1}{6}(1+\delta C_{Q})
J^{\alpha\beta\gamma\delta}
\nabla^{\mu}R_{\alpha\beta\gamma\delta},
\end{align}
\end{subequations}
respectively. The Christoffel connection in Eq.\,(\ref{Gamma-force}) is calculated from metric, including the SO, SS, and MQ effects. The constants $\delta C_{\vartheta}$ and $\delta C_{Q}$ are determined in the last two subsections \ref{subsec:NZscalar} and \ref{subsec:NZtensor}. Calculating the above geometric quantities up to the required order, we can write down the explicit expression of the EOM at 2PN approximation. It should be noted that, when deriving the EOM, the self field is removed via Hadamard regularization \cite{Yagi2012pn,Blanchet2014}, which is briefly introduced in Appendix \ref{appendix-hadamada}. We labeled these two BHs with $1$ and $2$. At the Newtonian order and leading dCS order, the motion of the $1$-st BH depends only on the mass, spin, and relative position of the $2$-nd BH, and vice versa. In terms of the 3-velocity, $\bm{v}_{1}$, the spatial components of 4-velocity, we can write down the acceleration of $1$-st BH \cite{Loutrel2018},
\begin{equation}
\label{a1-total}
\frac{dv_1^k}{dt}=a^{1,k}_{(\Gamma)}
+a^{1,k}_{\rm(SO)}
+a^{1,k}_{\rm(DD)}+a^{1,k}_{\rm(MQ)}.
\end{equation}
Each term in Eq.\,(\ref{a1-total}) can be divided into two parts: the GR part, $\bar{a}^{1,k}_{(\cdots)}$, and the part from dCS theory, $\delta a^{1,k}_{(\cdots)}$. Explicitly, they are expressed as
\begin{equation}
\label{a1-GR-Gamma}
\bar{a}^{1,k}_{(\Gamma)}
=-(\bar{\Gamma}_{1,00}^{k}
-2v_1^j\bar{\Gamma}_{1,0j}^{k}
+v_{1}^iv_{1}^j\bar{\Gamma}_{1,ij}^{k})
+v_{1}^k(\bar{\Gamma}_{1,00}^{0}
-2v_{1}^j\bar{\Gamma}_{1,0j}^{0}
+v_{1}^iv_{1}^j\bar{\Gamma}_{1,ij}^{0}),
\end{equation}
\begin{equation}
\label{a1-dCS-Gamma}
\delta a^{1,k}_{(\Gamma)}
=\partial_{k}\delta U_{1},
\end{equation}
\begin{equation}
\label{a1-GR-SO}
\bar{a}^{1,k}_{\rm(SO)}=\frac{1}{2m_{1}}\epsilon_{ijl}S_1^l
(2v_1^{j}\bar{R}_{1}^{i0k0}+\bar{R}_{1}^{ijk0}-v_{1}^{q}\bar{R}_{1}^{ijkq}),
\end{equation}
\begin{equation}
\label{a1-dCS-SO}
\delta a^{1,k}_{\rm(SO)}=0,
\end{equation}
\begin{equation}
\label{a1-GR-DD}
\bar{a}^{1,k}_{\rm(DD)}=0,
\end{equation}
\begin{equation}
\label{a1-dCS-DD}
\delta a^{1,k}_{\rm(DD)}
=\frac{1}{m_{1}^3}\delta C_{\vartheta}
S_{1}^{j}(\partial_{k}\partial_{j}\vartheta_{1}),
\end{equation}
\begin{equation}
\label{a1-GR-MQ}
\bar{a}^{1,k}_{\rm(MQ)}=-\frac{1}{2m_1^2}
(S_1^{i}S_1^{j}-S_1^2\delta_{ij})
(\partial_k\partial_i\partial_j\bar{U}_1),
\end{equation}
\begin{equation}
\label{a1-dCS-MQ}
\delta a^{1,k}_{\rm(MQ)}=-\frac{1}{2m_1^2}\delta C_{Q}
(S_1^{i}S_1^{j}-S_1^2\delta_{ij})
(\partial_k\partial_i\partial_j\bar{U}_1),
\end{equation}
where $\bar{\Gamma}_{1,\mu\nu}^{\lambda}$, $\bar{R}_{\mu\nu\rho\lambda}^{1}$, and $\bar{U}_{1}$ are the GR parts of the Christoffel connection, Riemann tensor, and Newtonian potential, Hadamard-regularized at the location of the $1$-st BH. In the above expressions, $\vartheta_{1}$ and $\delta U_{1}$ are Hadamard-regularized near-zone scalar field and modified Newtonian potential around the $1$-st BH.

For the GR case, all of the non-spinning effects (0PN, 1PN, and 2PN) are included in Eq.\,(\ref{a1-GR-SO}). The SO coupling, at linear spin order (\ref{a1-GR-Gamma}), is a 1.5PN effect. The SS and MQ couplings, at quadratic-spin order, are both 2PN effects. The former is included in (\ref{a1-GR-DD}), while the latter is included in Eqs.\,(\ref{a1-GR-Gamma}) and (\ref{a1-GR-MQ}). In particular, the dCS theory does not contribute to the SO coupling in the BBH system. The DD interaction in Eq.\,(\ref{a1-dCS-DD}) is proportional to $S_{1}S_{2}$, which is the modification of the SS coupling. In conclusion, the lowest-order modification induced by the dCS theory is at the 2PN order.

From the leading-order modified Newtonian potential (\ref{delta-U-BH-binary-reference}) and near-zone scalar field (\ref{scalar-near-zone-summation}), the Hadamard-regularized values of $\bar{U}_{1}$, $\vartheta_{1}$, and $\delta U_{1}$ at the position of the $1$-st BH are given by Appendix \ref{appendix-hadamada},
\begin{equation}
\label{U2-theta2-deltaU2}
\bar{U}_{1}=\frac{m_2}{r},
\quad
\vartheta_{1}=\frac{\delta C_{\vartheta}}{4\pi\beta}
\frac{S_{2}^{k}}{m_{2}^2}\frac{\hat{n}^{k}}{r^2},
\quad\text{and}\quad
\delta U_{1}
=\frac{603}{3584}\zeta\frac{m}{m_2}
\gamma^3\frac{1}{m_2^4}
\left[(\hat{\bm{n}}\cdot\bm{S}_{2})^2-\frac{1}{3}S_2^2\right].
\end{equation}
$r\equiv|\mathbf{z}_{1}-\mathbf{z}_2|$ is the relative distance (orbital radius) of the BBH, and $\hat{\bm{n}}\equiv\hat{\bm{n}}_{1}-\hat{\bm{n}}_{2}$ is the unit relative position vector. Additionally, we define the total mass of the binary system $m\equiv m_1+m_2$ and $\gamma\equiv m/r$. The dimensionless coupling normalized by the total mass is
\begin{equation}
\label{zeta}
\zeta\equiv 16\pi\frac{\alpha^2}{\beta m^4}.
\end{equation}
Substituting Eq.\,(\ref{U2-theta2-deltaU2}) into (\ref{a1-total})-(\ref{a1-dCS-MQ}), the dCS modification to the acceleration is thus calculated explicitly as
\begin{equation}
\label{delta-a1-total}
\begin{aligned}
\delta\bm{a}^{1}&=
\delta\bm{a}^{1}_{(\Gamma)}
+\delta\bm{a}^{1}_{\rm(DD)}
+\delta\bm{a}^{1}_{\rm(MQ)}\\
&=\gamma^4\zeta
\left\{\frac{603}{3584}\left[\frac{1}{m_2}\frac{1}{m_2^4}
\left[2(\hat{\bm{n}}\cdot\bm{S}_{2})\bm{S}_{2}
-5(\hat{\bm{n}}\cdot\bm{S}_{2})^2\hat{\bm{n}}
+S_2^2\hat{\bm{n}}\right]
+\frac{m_2}{m_1^2}\frac{1}{m_1^4}
\left[2(\hat{\bm{n}}\cdot\bm{S}_1)\bm{S}_1
-5(\hat{\bm{n}}\cdot\bm{S}_1)^2\hat{\bm{n}}
+S_1^2\hat{\bm{n}}\right]\right]\right.\\
&\qquad\qquad\left.-\frac{75}{256}
\frac{1}{m_1}\frac{1}{m_1^2}\frac{1}{m_{2}^2}
\left[(\bm{S}_1\cdot\bm{S}_2)\hat{\bm{n}}
+(\hat{\bm{n}}\cdot\bm{S}_1)\bm{S}_{2}
+(\hat{\bm{n}}\cdot\bm{S}_2)\bm{S}_1
-5(\hat{\bm{n}}\cdot\bm{S}_1)(\hat{\bm{n}}\cdot\bm{S}_2)
\hat{\bm{n}}\right]\right\}.
\end{aligned}
\end{equation}
This modification (\ref{delta-a1-total}) is very similar to the SS and MQ terms in GR, which are shown in Refs.\,\cite{ThorneHartle1985,Kidder1995}. Under Newton's gravity, SAM does not influence the motion of objects. However, in GR and its modification, SAM will produce the components of gravitational force along the SAM directions.

\subsection{Relative EOM for Spin-Aligned Quasi-Circular Orbits}
\label{subsec:relativeEOM}
Now, let us focus on the spin-aligned quasi-circular orbits, in which the SAM is perpendicular to the orbital plane, and the SAM of each object is conserved. The spin-aligned assumption means $\hat{\bm{n}}\cdot\bm{S}_{1}=\hat{\bm{n}}\cdot\bm{S}_{2}=0$. At the same time, when the energy dissipation is not taken into consideration, the relative orbit of the binary BHs is assumed to be circular, indicating that $\hat{\bm{n}}\cdot\bm{v}=0$, where $\bm{v}$ is the relative velocity, defined as $\bm{v}\equiv\bm{v}_1-\bm{v}_2$. The total acceleration of $1$-st BH (\ref{delta-a1-total}) thus becomes
\begin{equation}
\label{delta-a1-spin-aligned}
\delta\bm{a}^{1}=\zeta\gamma^4
\Bigg\{-\frac{75}{256}\frac{1}{m_1}
\left(\frac{\bm{S}_1}{m_1^2}
\cdot\frac{\bm{S}_2}{m_{2}^2}\right)\\
+\frac{603}{3584}\left[
\frac{m_2}{m_1^2}\left(\frac{\bm{S}_1}{m_1^2}\right)^2
+\frac{1}{m_2}\left(\frac{\bm{S}_2}{m_2^2}\right)^2\right]
\Bigg\}\hat{\bm{n}}.
\end{equation}
Exchanging the label $1$ and $2$ in Eq.\,(\ref{delta-a1-spin-aligned}), we obtain the EOM of $2$-nd BH, governed by the gravitational field of $1$-st BH, $\delta\bm{a}^2$. The relative acceleration $\delta\bm{a}$ is the difference between $\delta\bm{a}^1$ and $\delta\bm{a}^2$, 
\begin{equation}
\label{delta-a-spin-aligned}
\delta\bm{a}\equiv
\delta\bm{a}^1-\delta\bm{a}^2
=\zeta\frac{\gamma^3}{m}
\Bigg\{-\frac{75}{256}\frac{1}{\nu}
\left(\frac{\bm{S}_1}{m_1^2}
\cdot\frac{\bm{S}_2}{m_2^2}\right)
+\frac{603}{3584}\left[\frac{m^2}{m_1^2}
\left(\frac{\bm{S}_1}{m_1^2}\right)^2
+\frac{m^2}{m_2^2}
\left(\frac{\bm{S}_2}{m_2^2}\right)^2\right]\Bigg\}
\hat{\bm{n}},
\end{equation}
where $\nu\equiv m_1m_2/m^2$ is the symmetric mass ratio. Combined with the GR part, the total acceleration is written as 
\begin{equation}
\label{a-spin-aligned}
\frac{d\bm{v}}{dt}
=-\frac{\gamma^2}{m}\left[1+\bar{\mathcal{A}}_{1}
+\bar{\mathcal{A}}_{1.5}
+(\bar{\mathcal{A}}_{2}+\delta\mathcal{A})
\right]\hat{\bm{n}},
\end{equation}
up to 2PN order. For spin-aligned quasi-circular orbits, the relative acceleration (\ref{delta-a-spin-aligned}) is completely along the line between the two BHs. The coefficients $\bar{\mathcal{A}}_1$, $\bar{\mathcal{A}}_{1.5}$, and $\bar{\mathcal{A}}_2$, are 1PN, 1.5PN, and 2PN coefficients of the GR part. These coefficients are \cite{Blanchet1995,Kidder1995,Faye2006,Buonanno2013}
\begin{equation}
\label{mathcal-A1}
\bar{\mathcal{A}}_{1}=\left(1+3\nu\right)v^2-2\left(2+\nu\right)\gamma,
\end{equation}
\begin{equation}
\label{mathcal-A15}
\bar{\mathcal{A}}_{1.5}=-\gamma(s+3\delta\sigma)v,
\end{equation}
and
\begin{equation}
\label{mathcal-A2}
\bar{\mathcal{A}}_{2}=\nu(3-4\nu)v^{4}
-\gamma\nu\left(\frac{13}{2}-2\nu\right)v^2
+\gamma^2\left(9+\frac{87}{4}\nu\right)
+3\gamma^2\nu\left(\frac{\bm{S}_{1}}{m_1^2}
\cdot\frac{\bm{S}_{2}}{m_2^2}\right)
+\frac{3}{2}\gamma^2
\left[\frac{m_1^2}{m^2}
\left(\frac{\bm{S}_1}{m_1^2}\right)^2
+\frac{m_2^2}{m^2}
\left(\frac{\bm{S}_2}{m_2^2}\right)^2\right].
\end{equation}
The total spin is $\bm{S}\equiv\bm{S}_1+\bm{S}_2$ and $s\equiv|\bm{S}|/m^2$. The mass difference is $\delta\equiv (m_1-m_2)/m$ and the spin difference is $\bm{\Sigma}\equiv m(\bm{S}_2/m_2-\bm{S}_1/m_1)$, with $\sigma\equiv|\bm{\Sigma}|/m^2$. As we have mentioned, the SO coupling is included in the 1.5PN correction (\ref{mathcal-A15}), and the SS and MQ effects are included in the 2PN correction (\ref{mathcal-A2}). The dCS modification
\begin{equation}
\label{mathcal-delta-A}
\begin{aligned}
\delta\mathcal{A}&=\gamma^2
\left\{\frac{75}{256}\nu\zeta_{12}
\left(\frac{\bm{S}_1}{m_1^2}
\cdot\frac{\bm{S}_2}{m_2^2}\right)
-\frac{603}{3584}\left[\zeta_1\frac{m_1^2}{m^2}
\left(\frac{\bm{S}_1}{m_1^2}\right)^2
+\zeta_2\frac{m_2^2}{m^2}
\left(\frac{\bm{S}_2}{m_2^2}\right)^2\right]\right\}\\
&=\zeta\gamma^2
\left\{\frac{75}{256}\frac{1}{\nu}
\left(\frac{\bm{S}_1}{m_1^2}
\cdot\frac{\bm{S}_2}{m_2^2}\right)
-\frac{603}{3584}\left[\frac{m^2}{m_1^2}
\left(\frac{\bm{S}_1}{m_1^2}\right)^2
+\frac{m^2}{m_2^2}
\left(\frac{\bm{S}_2}{m_2^2}\right)^2\right]\right\}
\end{aligned}
\end{equation}
is also 2PN approximation. As shown in above equations, the dCS modification $\delta\mathcal{A}$ (\ref{mathcal-delta-A}) has similar dependence on spin with the quadratic-spin terms of  $\bar{\mathcal{A}}_{2}$ (\ref{mathcal-A2}). To obtain the dCS modification, we just need to take the replacements $\bm{S}_{1}\cdot\bm{S}_{2}\rightarrow(25/256)\zeta_{12}(\bm{S}_{1}\cdot\bm{S}_{2})$ for DD coupling and $\bm{S}_{A}^2\rightarrow-(201/1792)\zeta_{A}\bm{S}_{A}^2$ for MQ coupling. The DD effect of dCS modification in Eq.\,(\ref{mathcal-delta-A}) may provide attraction or repulsion. The scalar fields of two BHs with parallel (antiparallel) SAM will attract (repulse) each other, just like two parallel (antiparallel) magnetic moments, and provide negative (positive) potential energy. The modified MQ effect in Eq.\,(\ref{mathcal-delta-A}) generates repulsive force rather than gravity because $\delta C_{Q}<0$, as shown in Eq.\,(\ref{delta-C-Q}).

\subsection{Modified Kepler's Third Law}
\label{subsec:Kepler}
In Newtonian gravity, the orbital frequency $\omega$ is given by EOM (\ref{a-spin-aligned}), i.e., $|a|=\omega^2r$. Following the standard textbook, we define dimensionless frequency $x\equiv(m\omega)^{2/3}$. The modified Kepler's third law is the relationship between dimensionless frequency and distance, $x$ and $\gamma$, which are given by
\begin{equation}
\label{Kepler}
x=\gamma\left[1+\frac{1}{3}\gamma\bar{\varpi}_{1}
+\frac{1}{3}\gamma^{1.5}\bar{\varpi}_{1.5}
+\frac{1}{3}\gamma^2\left(-\frac{1}{3}\bar{\varpi}_{1}^2
+\bar{\varpi}_{2}+\delta\varpi\right)\right],
\end{equation}
\begin{equation}
\label{Kepler-inverse}
\gamma=x\left[1-\frac{1}{3}x\bar{\varpi}_{1}
-\frac{1}{3}x^{1.5}\bar{\varpi}_{1.5}
+\frac{1}{3}x^2\left(\bar{\varpi}_{1}^2
-\bar{\varpi}_{2}-\delta\varpi\right)\right].
\end{equation}
A set of higher PN coefficients are
\begin{equation}
\label{varpi1}
\bar{\varpi}_{1}=-3+\nu,
\end{equation}
\begin{equation}
\label{varpi15}
\bar{\varpi}_{1.5}
=-5s-3\delta\sigma,
\end{equation}
\begin{equation}
\label{varpi2}
\bar{\varpi}_{2}
=6+\frac{41}{4}\nu+\nu^2+3\nu
\left(\frac{\bm{S}_{1}}{m_1^2}\cdot\frac{\bm{S}_{2}}{m_2^2}\right)
+\frac{3}{2}
\left[\frac{m_1^2}{m^2}
\left(\frac{\bm{S}_1}{m_1^2}\right)^2
+\frac{m_2^2}{m^2}
\left(\frac{\bm{S}_2}{m_2^2}\right)^2\right],
\end{equation}
and
\begin{equation}
\label{delta-varpi}
\delta\varpi=\zeta
\Bigg\{\frac{75}{256}\frac{1}{\nu}
\left(\frac{\bm{S}_1}{m_1^2}\cdot\frac{\bm{S}_2}{m_2^2}\right)
-\frac{603}{3584}\left[\frac{m^2}{m_1^2}
\left(\frac{\bm{S}_1}{m_1^2}\right)^2
+\frac{m^2}{m_2^2}
\left(\frac{\bm{S}_2}{m_2^2}\right)^2\right]\Bigg\}.
\end{equation}
Combined with frequency evolution, the modified Kepler's third law can be used to calculate the parameterized post-Einsteinian (ppE) parameters without calculating the full waveform. In terms of $\gamma$ and Eqs.\,(\ref{varpi1})-(\ref{delta-varpi}), the relative EOM of the binary system (\ref{a-spin-aligned}) is rewritten as
\begin{equation}
\label{a-spin-aligned-kepler}
\frac{d\bm{v}}{dt}
=-\frac{\gamma^2}{m}\left[1+\gamma\bar{\varpi}_{1}
+\gamma^{1.5}\bar{\varpi}_{1.5}
+\gamma^2(\bar{\varpi}_{2}+\delta\varpi)\right]\hat{\bm{n}},
\end{equation}
up to the 2PN order. The EOM is also can re-expressed in terms of symbol dimensionless frequency $x$.

\subsection{Conserved Energy}
\label{subsec:energy}
{\color{black}From the EOM (\ref{a-spin-aligned-kepler}), the conserved energy of the binary system can be given by ``guess-work" \cite{Faye2006,Will2014}. We know that at the Newtonian limit, the conserved energy is $E=(1/2)\mu v^2-\mu\gamma$. All the possible contributions of conserved energy from dCS modification must be proportional to $\bm{S}_{1}\cdot\bm{S}_{2}$, $(\hat{\bm{n}}\cdot\bm{S}_{1})(\hat{\bm{n}}\cdot\bm{S}_{2})$, $S_{1}^2$, $S_{2}^2$, $(\hat{\bm{n}}\cdot\bm{S}_{1})^2$, and $(\hat{\bm{n}}\cdot\bm{S}_{2})^2$. Then the correct combination of such terms can be identified by including them all (with unknown coefficients) in a trial expression for $E$ and demanding that $dE/dt=0$ by virtue of the full 2PN EOM. The binding energy of the BBH system with spin-aligned quasi-circular orbits is given by}
\begin{equation}
\label{binding-energy}
E(x)=-\frac{\mu x}{2}\left[1
+x\bar{\varepsilon}_{1}
+x^{1.5}\bar{\varepsilon}_{1.5}
+x^2(\bar{\varepsilon}_{2}
+\delta\varepsilon)\right],
\end{equation}
with reduced mass being defined as $\mu\equiv m_1m_2/m=\nu m$. The higher-order coefficients in Eq.\,(\ref{binding-energy}) are
\begin{equation}
\label{epsilon1}
\bar{\varepsilon}_{1}=-\frac{3}{4}-\frac{1}{12}\nu,
\end{equation}
\begin{equation}
\label{epsilon15}
\bar{\varepsilon}_{1.5}=
\frac{14}{3}s
+2\delta\sigma,
\end{equation}
\begin{equation}
\label{epsilon2}
\bar{\varepsilon}_{2}=-\frac{27}{8}+\frac{19}{8}\nu-\frac{1}{24}\nu^2
-2\nu\left(\frac{\bm{S}_{1}}{m_1^2}\cdot\frac{\bm{S}_{2}}{m_2^2}\right)
-\left[\frac{m_1^2}{m^2}\left(\frac{\bm{S}_1}{m_1^2}\right)^2
+\frac{m_2^2}{m^2}\left(\frac{\bm{S}_2}{m_2^2}\right)^2\right],
\end{equation}
and
\begin{equation}
\label{delta-epsilon}
\delta\varepsilon=\zeta
\left\{-\frac{25}{128}\frac{1}{\nu}
\left(\frac{\bm{S}_1}{m_1^2}\cdot\frac{\bm{S}_2}{m_2^2}\right)
+\frac{201}{1792}
\left[\frac{m^2}{m_1^2}
\left(\frac{\bm{S}_1}{m_1^2}\right)^2
+\frac{m^2}{m_2^2}
\left(\frac{\bm{S}_2}{m_2^2}\right)^2
\right]
\right\}.
\end{equation}
As we have derived, the DD interaction potential is positive (negative) for antiparallel (parallel) SAM binary BHs and the MQ potential is positive for both parallel and antiparallel cases. The conserved energy provides the energy of gravitational radiation. This result is consistent with that in Refs.\,\cite{Yagi2012gw,Yagi2016e}, in which the scalar fields are seen as magnetic dipoles, and the modified Dixon tensor is seen as the effective density of BH. In the simplest case, where the $1$-st and $2$-nd BHs have roughly the same mass and SAM, the ratio of binding energy caused by the SS and MQ effect is thus about $0.87$. In conclusion, the correction of SS and MQ effect on binding energy is comparable.

{\color{black}The acceleration (\ref{mathcal-delta-A}), modified Kepler's third law (\ref{Kepler}) and binding energy (\ref{binding-energy}), are consistent with that in Refs.\,\cite{Yagi2012gw,Tahura2018}, although completely different methods are adopted. Our method is based on the modified MPD equations in dCS theory and guess-work provided by Refs.\,\cite{Will2014,Faye2006}, which generally describe the motion and precession of a spinning particle. In Ref.\,\cite{Yagi2012gw}, the modified Newtonian potential satisfies the Poisson equation, $\bm{\nabla}^2\delta U=4\pi(\delta\rho_{1}+\delta\rho_{2})$, with $\delta\rho_{A}$ being the effective density induced by the dCS modification of the $A$-th BH. Then, the potential energy density from modified MQ coupling is determined by $\delta\rho_{1}U_{2}+\delta\rho_{2}U_{1}+\rho_{1}\delta U_{2}+\rho_{2}\delta U_{1}$, with $\rho_{A}$ being the mass density of the $A$-th BH. At the same time, the binding energy between the scalar fields of each BH is calculated by regarding them as two magnetic dipoles. The interaction force is defined as the derivative of effective potential energy with respect to the distance between BBH. The approach in Ref.\,\cite{Yagi2012gw} is reasonable up to the lowest-order PN correction. But our method is more general, especially for the higher accuracy template calculation in the future.}

\section{Radiation Field}
\label{sec:radiation}
% In this section, we calculate the radiation field at the wave zone by multipole moment formula. The scalar waveform is given in subsection \ref{subsec:scalarradiation}. Subsection \ref{subsec:tensorradiation} reviews the 2PN waveform in GR and calculates the deformation waveform. The orbital phase and the waveform at propagation frame is given in subsection \ref{subsec:propagation-frame}.

\subsection{Scalar Radiation}
\label{subsec:scalarradiation}
The scalar radiation is obtained from the scalar perturbation equation (\ref{scalar-perturbation-modification}), $\Box^2_{\eta}\vartheta=-16\pi\sigma+4\pi\rho_{\vartheta}$, where $\sigma\equiv\sigma_1+\sigma_2$ and $\rho_{\vartheta}$ are the non-compact and compact sources of the scalar field. In terms of the near-zone solution of the scalar field (\ref{scalar-near-zone-summation}) and the PN metric (\ref{hh-near-zone}), the source terms $\sigma_1$ and $\sigma_2$ are written as
\begin{equation}
\label{sigma}
16\pi\sigma_{1}
=-2\bar{U}(\partial_{i}\partial_{i}\vartheta^{\rm(B)})
\sim\mathcal{O}(1/c^6),
\quad\text{and}\quad
16\pi\sigma_{2}
=\frac{8\alpha}{\beta}\epsilon^{ijk}
(\partial_{i}\partial_{m}\bar{U})
(\partial_{j}\partial_{m}\bar{V}_{k})
\sim\mathcal{O}(1/c^9).
\end{equation}
respectively. The monopole, dipole, and quadrupole radiation \cite{Maggiore2008} generated by non-compact sources are
\begin{subequations}
\label{theta-sigma-def}
\begin{align}
\label{theta-sigma-mon-def}
\vartheta^{(\sigma)}_{\rm mon}
&=\frac{4}{R}\int\sigma(t_r,\mathbf{x}')d^3\mathbf{x}',\\
\label{theta-sigma-dip-def}
\vartheta^{(\sigma)}_{\rm dip}
&=\frac{4}{R}\frac{\partial}{\partial t}\int\sigma(t_r,\mathbf{x}')
(\hat{\mathbf{N}}\cdot\mathbf{x}')d^3\mathbf{x}',\\
\label{theta-sigma-quad-def}
\vartheta^{(\sigma)}_{\rm quad}
&=\frac{4}{R}\frac{1}{2}\frac{\partial^2}{\partial t^2}\int\sigma(t_r,\mathbf{x}')
(\hat{\mathbf{N}}\cdot\mathbf{x}')^{2}d^3\mathbf{x}'.
\end{align}
\end{subequations}
In Eq.\,(\ref{theta-sigma-def}), $R$ is the distance between the observer and the GW source system \footnote{Please distinguish the distance $R$ and the Ricci scalar $R$ according to the situation.}, $t_{r}\equiv t-R$ is retarded time, and $\hat{\mathbf{N}}$ is the GW propagation direction. The source $\sigma$ (\ref{sigma}) is the product of two near-zone solutions, e.g., $\bar{U}\vartheta^{\rm(B)}$ and $\bar{U}\bar{V}_k$, etc. Every solution is written in the form, e.g., $\bar{U}=\bar{U}_1+\bar{U}_2$, $\bar{V}^{k}=\bar{V}_1^{k}+\bar{V}_2^{k}$, etc. Thus, the retarded integration consists of the self-interaction terms like $\bar{U}_1\bar{V}_1^k$ and cross-interaction terms like $\bar{U}_1\bar{V}_2^k$ (\ref{theta-sigma-def}). The full volume integration of $1/r_1^2$ is divergent, without physical meaning. The subsequent calculation will ignore the self-interaction term. The integration of cross-interaction terms like $1/(r_1r_2)$ can be managed appropriately by Hadamard regularization \cite{Yagi2012pn,Blanchet2014},
\begin{equation}
\label{Hadamard}
\int\frac{d^3\mathbf{x}}{r_{1}r_{2}}=-2\pi r.
\end{equation}
The dominant contribution comes from the monopole term (\ref{theta-sigma-mon-def}), given by
\begin{equation}
\label{theta-sigma-mon}
\vartheta^{(\sigma)}_{\rm mon}
=-\frac{2}{R}\frac{m}{r^2}\left[
\frac{m_2}{m}(\hat{\bm{n}}\cdot\bm{\mu}_{1})
-\frac{m_1}{m}(\hat{\bm{n}}\cdot\bm{\mu}_{2})\right].
\end{equation}
The dipole (\ref{theta-sigma-dip-def}) and quadrupole radiation (\ref{theta-sigma-quad-def}) enter higher-order PN approximation \cite{Yagi2012pn,Yagi2016e}. The estimations are shown in Appendix \ref{appendix-scalar}.

Similarly, the multipole moment formula gives the radiation by the source with compact support,
\begin{subequations}
\label{theta-rho-def}
\begin{align}
\label{theta-rho-mon-def}
\vartheta^{(\rho)}_{\rm mon}
&=-\frac{1}{R}\int\rho_{\vartheta}(t_r,\mathbf{x}')d^3\mathbf{x}',\\
\label{theta-rho-dip-def}
\vartheta^{(\rho)}_{\rm dip}
&=-\frac{1}{R}\frac{\partial}{\partial t}
\int \rho_{\vartheta}(t_r,\mathbf{x}')
(\hat{\mathbf{N}}\cdot\mathbf{x}')d^3\mathbf{x}',\\
\label{theta-rho-quad-def}
\vartheta^{(\rho)}_{\rm quad}
&=-\frac{1}{R}\frac{1}{2}\frac{\partial^2}{\partial t^2}
\int \rho_{\vartheta}(t_r,\mathbf{x}')
(\hat{\mathbf{N}}\cdot\mathbf{x}')^{2}d^3\mathbf{x}'.
\end{align}
\end{subequations}
After calculating Eq.\,(\ref{theta-rho-def}), the dominant terms are
\begin{equation}
\label{theta-rho-mon}
\vartheta^{(\rho)}_{\rm mon}
=\frac{1}{R}\frac{m}{r^2}
\left[\frac{m_{2}}{m}(\hat{\bm{n}}\cdot\bm{\mu}_{1})
-\frac{m_{1}}{m}(\hat{\bm{n}}\cdot\bm{\mu}_{2})\right],
\end{equation}
and 
\begin{equation}
\label{theta-rho-quad}
\vartheta^{(\rho)}_{\rm quad}
=-\frac{1}{R}\frac{m}{r^2}
\left[\frac{m_{2}}{m}(\hat{\mathbf{N}}\cdot\bm{\mu}_{1})
-\frac{m_{1}}{m}(\hat{\mathbf{N}}\cdot\bm{\mu}_{2})\right]
(\hat{\mathbf{N}}\cdot\hat{\bm{n}}).
\end{equation}
The dipole radiation (\ref{theta-rho-dip-def}) enters higher PN order. The estimations are also listed in Appendix \ref{appendix-scalar}.

As shown in Eqs.\,(\ref{scalar-compact-source-123}, \ref{sigma}), the magnitudes of the scalar source terms are $\sigma_{1}\sim\mathcal{O}(1/c^6)$,
$\sigma_{2}\sim\mathcal{O}(1/c^9)$, 
$\rho_{1}\sim\mathcal{O}(1/c^4)$, 
$\rho_{2}\sim\mathcal{O}(1/c^6)$, and $\rho_{3}\sim\mathcal{O}(1/c^6)$, respectively.
Generally, the dominant contribution is sourced by $\rho_{1}$. However, the monopole and dipole radiations of $\rho_1$ vanish because the spin-aligned assumption and that dCS theory requires the scalar field to be dipole (see Eqs.\,(\ref{theta-rho1-mon}, \ref{theta-rho1-dip})) \cite{Yagi2012}. Such that, the monopole radiation of $\sigma_1$, the quadrupole radiation of $\rho_{1}$, and the monopole radiation of $\rho_2$ are of order $1/c^6$, while the monopole radiation of $\rho_3$ vanishes due to dipole scalar field. These terms contribute to the dominant scalar radiation, which are $1/c^2$ higher than leading-order gravitational radiation. The leading-order dCS scalar radiation enters 2PN modification in energy flux relative to GW radiation (of order $\mathcal{O}(1/c^4)$).

Let's make a comparison between dCS theory and BD theory. In the BD gravity, the scalar field $\varphi$ satisfies equation $\Box_{\eta}^2\varphi=-16\pi\rho(1-2s)/(4\omega_0+6)$, with $s$ being the sensitivity of objects and $\omega_0$ being the coupling constant. The matter energy density is of order $1/c^2$. The monopole scalar radiation vanishes as shown in \cite{XingZhang2017BD,TanLiu2020BD}. The leading-order (dipole) scalar radiation is of order $1/c^3$, which is $1/c$ lower than leading-order gravitational radiation. Finally, the leading-order BD scalar radiation enters -1PN modification in energy flux relative to GW radiation.

Combined Eqs.\,(\ref{theta-sigma-mon}), (\ref{theta-rho-mon}), and (\ref{theta-rho-quad}), the total quadrupole scalar radiation is written as
\begin{equation}
\label{theta-radiation}
\vartheta
=\frac{5}{8}\frac{\mu}{R}\frac{\alpha}{\beta m^2}\frac{\gamma^2}{\nu}
(\hat{\mathbf{N}}\cdot\bm{\Delta})
(\hat{\mathbf{N}}\cdot\hat{\bm{n}})
\sim\mathcal{O}(1/c^6),
\quad\text{with}\quad
\bm{\Delta}\equiv
\frac{m_{2}}{m}\frac{\bm{S}_{1}}{m_1^2}
-\frac{m_{1}}{m}\frac{\bm{S}_{2}}{m_2^2}.
\end{equation}
This result (\ref{theta-radiation}) will be used to calculate the energy flux carried by gravitational radiation. Comparing Eq.\,(\ref{theta-radiation}) with the previous results given in Refs.\,\cite{Yagi2012pn,Yagi2016e}, some terms are omitted because of the spin-aligned assumption.

\subsection{Tensor Radiation}
\label{subsec:tensorradiation}
In the PN framework, the gravitational waveform contains an ``instantaneous" term, depending on the state of the binary at the retarded time $t_r$ only, and a ``tail" term, which is sensitive to the wave field at all previous time $t_r-\tau$. We write the waveform as two pieces,
\begin{equation}
\label{H-inst-tail}
\bar{H}_{ij}
=(\bar{H}_{ij})_{\rm inst}+(\bar{H}_{ij})_{\rm tail}.
\end{equation}
The instantaneous and tail terms are given by \cite{Blanchet1995,Blanchet2014} 
\begin{equation}
\label{H-inst-def}
\begin{aligned}
(\bar{H}_{ij})_{\rm inst}
&=\frac{2}{R}\Bigg\{\bar{\mathcal{I}}^{(2)}_{ij}
+\left[\frac{1}{3}\bar{\mathcal{I}}^{(3)}_{ija}\hat{N}^{a}
+\frac{4}{3}\epsilon_{ab(i}\bar{\mathcal{J}}^{(2)}_{j)a}\hat{N}^{b}\right]
+\left[\frac{1}{12}\bar{\mathcal{I}}^{(4)}_{ijab}\hat{N}^{ab}
+\frac{1}{2}\epsilon_{ab(i}\bar{\mathcal{J}}^{(3)}_{j)ac}\hat{N}^{bc}\right]\\
&+\left[\frac{1}{60}\bar{\mathcal{I}}^{(5)}_{ijabc}\hat{N}^{abc}
+\frac{2}{15}\epsilon_{ab(i}\bar{\mathcal{J}}_{j)acd}\hat{N}^{bcd}\right]
+\left[\frac{1}{360}\bar{\mathcal{I}}^{(6)}_{ijabcd}\hat{N}^{abcd}
+\frac{1}{36}\epsilon_{ab(i}\bar{\mathcal{J}}^{(5)}_{j)acde}\hat{N}^{bcde}\right]
+\delta{\mathcal{I}}^{(2)}_{ij}\Bigg\},
\end{aligned}
\end{equation}
and
\begin{equation}
\label{H-tail-def}
\begin{aligned}
(\bar{H}_{ij})_{\rm tail}=\frac{4m}{R}\int_{0}^{+\infty}
d\tau\left[
\ln\left(\frac{\tau}{\tau_1}\right)
\left(\bar{\mathcal{I}}_{ij}+\delta\mathcal{I}_{ij}\right)^{(4)}
+\frac{1}{3}\ln\left(\frac{\tau}{\tau_2}\right)
\hat{N}_{a}\bar{\mathcal{I}}_{ija}^{(5)}
+\frac{4}{3}\ln\left(\frac{\tau}{\tau_3}\right)\epsilon_{ab(i}\hat{N}_{b}
\bar{\mathcal{J}}_{j)a}^{(4)}\right](t_r-\tau),
\end{aligned}
\end{equation}
with $\hat{N}^{i_1i_2\cdots i_l}\equiv\hat{N}^{i_1}\hat{N}^{i_2}\cdots\hat{N}^{i_l}$. $\bar{\mathcal{I}}_{L}$ and $\bar{\mathcal{J}}_{iL}$ are mass and current multipole moments in GR. $\tau_1, \tau_2, \tau_3$ are arbitrary constants \cite{Blanchet1995}. All of these moments can be divided into mass (M), spin (S), and quadratic-spin (Q) parts,
\begin{equation}
\label{mass-current-def}
\bar{\mathcal{I}}_{L}
=\bar{\mathcal{I}}_{L}^{\rm(M)}
+\bar{\mathcal{I}}_{L}^{\rm(S)}
+\bar{\mathcal{I}}_{L}^{\rm(Q)},\quad
\bar{\mathcal{J}}_{iL}
=\bar{\mathcal{J}}_{iL}^{\rm(M)}
+\bar{\mathcal{J}}_{iL}^{\rm(S)}
+\bar{\mathcal{J}}_{iL}^{\rm(Q)}.
\end{equation}
For example, the mass quadrupole moment in GR is \cite{Blanchet1995,Kidder1995,Faye2006,Bohe2015,Buonanno2013}
\begin{equation}
\label{mass-quad-M}
\begin{aligned}
\bar{\mathcal{I}}_{ij}^{\rm(M)}&=\mu r^2\Bigg\{
\left[1+\gamma\left(-\frac{1}{42}-\frac{13}{14}\nu\right)
+\gamma^2\left(-\frac{461}{1512}
-\frac{18395}{1512}\nu
-\frac{241}{1512}\nu^2\right)\right]
(\hat{n}^i\hat{n}^j)^{\rm STF}\\
&\qquad+\left[\left(\frac{11}{21}-\frac{11}{7}\nu\right)
+\gamma\left(\frac{1607}{378}
-\frac{1681}{378}\nu+\frac{229}{378}\nu^2\right)\right]
(v^iv^j)^{\rm STF}\Bigg\},
\end{aligned}
\end{equation}
\begin{equation}
\label{mass-quad-S}
\bar{\mathcal{I}}_{ij}^{\rm(S)}
=\nu r\left\{
\frac{8}{3}[\hat{n}_i(\bm{v}\times\bm{S})_{j}]^{\rm STF}
-\frac{4}{3}[v_i(\hat{\bm{n}}\times\bm{S})_{j}]^{\rm STF}
+\frac{8}{3}\frac{\delta m}{m}
[\hat{n}_i(\bm{v}\times\bm{\Sigma})_{j}]^{\rm STF}
-\frac{4}{3}\frac{\delta m}{m}
[v_i(\hat{\bm{n}}\times\bm{\Sigma})_{j}]^{\rm STF}\right\},
\end{equation}
and
\begin{equation}
\label{mass-quad-Q}
\bar{\mathcal{I}}_{ij}^{\rm(Q)}
=-\left[\frac{1}{m_1}(S_1^iS_1^j)^{\rm STF}
+\frac{1}{m_2}(S_2^iS_2^j)^{\rm STF}\right].
\end{equation}
STF means symmetric tracefree tensor, e.g., $(\hat{n}_i\hat{n}_j)^{\rm STF}=\hat{n}_i\hat{n}_j-(1/3)\delta_{ij}$ \cite{Maggiore2008}. The higher-order mass and current moments can be found in Refs.\,\cite{Blanchet1995,Blanchet2006,Buonanno2013,Bohe2015}. For spin-aligned quasi-circular orbits, the quadratic-spin part (\ref{mass-quad-Q}) is conserved. Thus they will not contribute to gravitational radiation. Similarly, the dCS modification of mass quadrupole moment,
\begin{equation}
\label{delta-mass-quad}
\delta\mathcal{I}_{ij}
=\frac{201}{1792}\zeta
\left[\frac{m^4}{m_1^5}(S_1^iS_1^j)^{\rm STF}
+\frac{m^4}{m_2^5}(S_2^iS_2^j)^{\rm STF}\right],
\end{equation}
is also conserved, i.e., $\delta\bar{\mathcal{I}}_{ij}^{(2)}=0$ and $\delta\bar{\mathcal{I}}_{ij}^{(4)}=0$. The dCS modification to the fourth-order time derivative of mass quadrupole moment enters higher order. Thus the tail term (\ref{H-tail-def}) has no dCS modification. The dCS modification only comes from the second-order time derivative of the mass quadrupole moment. The total gravitational radiation waveform is \cite{Blanchet1995,Arun2009}
\begin{equation}
\label{Hij-waveform}
\begin{aligned}
\bar{H}_{ij}
&=\frac{2\mu}{R}\Big[
2(v_iv_j-\gamma\hat{n}_i\hat{n}_j)
+\{\text{Higher-Order PN Waveforms in GR}\}\Big]+\bar{k}_{ij}.
\end{aligned}
\end{equation}
We list the leading-order and dCS waveform in Eq.\,(\ref{Hij-waveform}) for simplification. The 0.5PN, $\cdots$, 2PN waveform in GR can be found in Refs.\,\cite{Blanchet1995,Kidder1995,Buonanno2013,Bohe2015,Blanchet2006}, including the SO effect at 1.5PN order, SS and MQ effects at 2PN order.

The metric-deformation radiation $\bar{k}_{ij}$ is explicitly expressed as
\begin{equation}
\label{kij}
\bar{k}_{ij}=-\frac{4\mu\gamma^3}{R}\delta\varpi
\hat{n}^i\hat{n}^j
\approx
-\frac{4\mu x^3}{R}\delta\varpi
\hat{n}^i\hat{n}^j.
\end{equation}
The coefficient $\delta\varpi$ is proportional to the dimensionless coupling constant $\zeta$, containing $\bm{S}_1^2$ -, $\bm{S}_2^2$ -, and $\bm{S}_1\cdot\bm{S}_2$ -terms, defined in Eqs.\,(\ref{Kepler}) and (\ref{delta-varpi}). The waveform (\ref{kij}) is similar to the quadratic-spin waveform in GR except for a coupling constant.

The complete metric-deformation waveform is not presented in Ref.\,\cite{Yagi2012pn} because some source terms in the tensor perturbation equation are missed. The erratum \cite{Yagi2016e} corrected the scalar and tensor perturbation equation, including all possible source terms, concluded that the energy flux carried by metric-deformation radiation is at 3PN order, far less than that of the scalar radiation (2PN). Such that the metric-deformation waveform is not given. However, we find that the scalar and tensor flux are both at 2PN order. The reasons are as follows. The lowest-order radiation is given by retarded-potential integral of the source and can be rewritten via conservation law, $\partial_{\mu}K^{\mu\nu}=0$,
\begin{equation}
\label{transform}
\bar{k}_{ij}\approx\frac{4}{R}\int K^{ij}d^3\mathbf{x}'
=\frac{2}{R}
\left[2\int\partial_{k}(x^{\prime i}K^{kj})d^3\mathbf{x}'
+\partial_{0}\int
\partial_{k}\left(x^{\prime i}x^{\prime j}K^{0k}\right)d^3\mathbf{x}'
+\partial_{0}\partial_{0}\int
x^{\prime i}x^{\prime j}K^{00}d^3\mathbf{x}'\right].
\end{equation}
The source $K_{ij}$ contains compact-support terms (the first two terms in Eq.\,(\ref{tensor-source})) and non-compact terms (the last three terms in Eq.\,(\ref{tensor-source})). The compact sources tend to $0$, strictly, while the non-compact sources tend to $1/R^2$ at far regions. For the compact case, the two complete-derivative integral in the square bracket of the Eq.\,(\ref{transform}) can be transformed into two infinity-sphere integral and then vanish. However, for the non-compact case, these two volume integral should be calculated as follow.
\begin{equation}
\begin{aligned}
\int\partial_{k}(x^{\prime i}K^{kj})d^3\mathbf{x}'
&=\oint(x^{\prime i}K^{kj})dS_{k}
=\oint\hat{N}^{k}(x^{\prime i}K^{kj})R^2d\Omega
\propto R\frac{1}{R^4}R^2=\frac{1}{R}\rightarrow0,\\
\int\partial_{k}(x^{\prime i}x^{\prime j}K^{0k})d^3\mathbf{x}'
&=\oint(x^{\prime i}x^{\prime j}K^{0k})dS_{k}
=\oint\hat{N}^{k}(x^{\prime i}x^{\prime j}K^{0k})R^2d\Omega
\propto R^2\frac{1}{R^4}R^2
\rightarrow\text{finite value}.
\end{aligned}
\end{equation}
According to the above discussion about gravitational radiation, we conclude that
\begin{equation}
\label{nonequal}
\frac{4}{R}\int K^{ij}d^3\mathbf{x}'
\neq\frac{2}{R}\partial_{0}\partial_{0}\int
x^{\prime i}x^{\prime j}K^{00}d^3\mathbf{x}'.
\end{equation}
Such that, the estimations to the PN order in the Eqs.\,(11, 16, 17) in Ref.\,\cite{Yagi2016e} are correct, which are 2PN corrections, but Eqs.\,(19-21) (in Ref.\,\cite{Yagi2016e}) are incorrect, because they assumed that the left-hand side equals to the right-hand side of Eq.\,(\ref{nonequal}) for the non-compact source. Therefore, we recalculate the full metric-deformation waveform in this section for spin-aligned case, the standard PN method \cite{Blanchet2014} is adopted. As shown in our result (\ref{kij}), the gravitational radiation contributes to the energy flux at 2PN order rather than 3PN as predicted by Ref.\,\cite{Yagi2016e}. Accordingly, the calculation of total radiant energy flux and ppE parameters in Refs.\,\cite{Yagi2016e,Yagi2012gw} are also inaccurate.

% It should be noted that since the source term, $K_{\mu\nu}$ (\ref{tensor-source}), is non-compact and does not tend to zero at infinity, the integral of $K_{ij}$ on the infinite sphere may also not be zero. Therefore, the retarded integration of $K_{ij}$ cannot be converted into the second-order time derivative of the integration of $K_{00}$, even though $K_{\mu\nu}$ is covariant-conserved, $\nabla_{\mu}K^{\mu\nu}=0$. 

Moreover, another equivalent derivation is shown in Appendix \ref{appendix-multipole}. The radiation field of the metric deformation is calculated through the monopole formula of $K_{ij}$ rather than the quadrupole formula of $K_{00}$ because of Eq.\,(\ref{nonequal}). Thus, the EOM of the BBH is not required for calculating the radiation in the wave zone. On the other hand, the time derivative of $K_{ij}$ is of higher PN order than we required, so it will not be included during calculation by the multipole moment formulas. However, the time derivative of the lowest-order moment will bring the modification of the EOM of the BBH. In the required order, the two methods are equivalent. Because the conservation of the matter EMT and the gravitational field determines the motion of the matter field \cite{Roshan2013}, The influence caused by the modified EOM on the lowest-order moment is equivalent to that caused by the modified source term on the higher-order moments.

\subsection{Propagation Frame}
\label{subsec:propagation-frame}
Following Ref.\,\cite{Maggiore2008}, we set the propagation direction of GWs as $\hat{\mathbf{N}}=(0,\sin\iota,\cos\iota)$, with the inclination angle $\iota$ between the orbital plane and propagation direction. The relative position and velocity of BBH are $\hat{\bm{n}}=(\cos\phi,\sin\phi,0)$ and $\bm{v}=\omega r(-\sin\phi,\cos\phi,0)$, where $\phi$ is the orbital phase evaluated at retarded time, defined as $\phi(t)\equiv\omega t_r=\omega(t-R)$. We then transform the metric deformation $\bar{k}_{ij}$ (\ref{kij}) from binary frame to propagation frame, in which propagation direction is regarded as the $z$-axis. Through the rotation matrix, 
\begin{equation}
\label{rotation-matrix}
\bm{\mathcal{R}}(\hat{\mathbf{N}})
=\left(\begin{array}{ccc}
1 & 0 & 0\\
0 & \cos\iota & -\sin\iota\\
0 & \sin\iota &  \cos\iota
\end{array}\right),
\end{equation}
the metric deformation $\bar{k}_{ij}$ (\ref{kij}) becomes
\begin{equation}
\label{k-propagation-frame}
\bar{k}_{ij}(\hat{\mathbf{N}})
=-\frac{2\mu x^3}{R}\delta\varpi
\left(\begin{array}{ccc}
\cos2\phi& \cos\iota\sin2\phi & \sin\iota\sin2\phi\\
\cos\iota\sin2\phi &-\cos^2\iota\cos2\phi&
-\sin\iota\cos\iota\cos2\phi\\
\sin\iota\sin2\phi
&-\sin\iota\cos\iota\cos2\phi&-\sin^2\iota\cos2\phi
\end{array}\right).
\end{equation}
Several constants have been omitted because they do not contribute to energy flux carried by radiation. This result (\ref{k-propagation-frame}) will be used to study the polarization modes and the energy flux carried by the tensor radiation.

\section{Polarization Modes}
\label{sec:polarization}
The polarization modes of GWs from BBH system with spin-aligned quasi-circular orbit in dCS gravity are studied through NP formalism \cite{NewmanPenrose1962}. The NP null tetrads are reviewed, and the polarization modes of metric deformation are obtained.
% in subsection \ref{subsec:nulltetrad}
% in subsection \ref{subsec:modes}

\subsection{Null Tetrad}
\label{subsec:nulltetrad}
The NP null tetrad is $\bm{e}_{(a)}\equiv(\bm{l},\bm{q},\bm{m},\bar{\bm{m}})$ \cite{Chandra1983,Wagle2019}, where $\bar{\bm{m}}$ is the complex conjugation of vector, $\bm{m}$ and the definitions of $\bm{l}$, $\bm{q}$, and $\bm{m}$ are
\begin{equation}
\label{lqm-def}
\bm{l}=\frac{1}{\sqrt{2}}(1,0,0,1),\quad
\bm{q}=\frac{1}{\sqrt{2}}(1,0,0,-1),\quad\text{and}\quad
\bm{m}=\frac{1}{\sqrt{2}}(0,1,i,0).
\end{equation}
These basis vectors are lightlike and satisfy the orthogonality, $\bm{l}\cdot\bm{l}=\bm{q}\cdot\bm{q}=\bm{m}\cdot\bm{m}=\bar{\bm{m}}\cdot\bar{\bm{m}}=0$ and $\bm{l}\cdot\bm{q}=-1$, $\bm{m}\cdot\bar{\bm{m}}=1$. The flat metric relates the null tetrads by $\eta^{\mu\nu}=-l^{\mu}q^{\nu}-q^{\mu}l^{\nu}+m^{\mu}\bar{m}^{\nu}+\bar{m}^{\mu}m^{\nu}$. Projecting this equation onto null tetrad $\bm{e}_{(a)}$, we obtain the flat metric in NP formalism \cite{Chandra1983},
\begin{equation}
\label{eta-def}
\eta_{(a)(b)}\equiv e_{(a)}^{\mu}e_{(b)}^{\nu}\eta_{\mu\nu}
=\left(\begin{array}{cccc}
0&-1&0&0\\
-1&0&0&0\\
0&0&0&1\\
0&0&1&0\\
\end{array}\right).
\end{equation}
$\eta_{\mu\nu}$ and $\eta^{\mu\nu}$ are used to raise and fall the coordinate indices, and $\eta_{(a)(b)}$ and $\eta^{(a)(b)}$ (\ref{eta-def}) to raise and fall the tetrad indices. All geometric quantities can be projected onto null tetrad $\bm{e}_{(a)}$, e.g., Riemann tensor
\begin{equation}
R_{(a)(b)(c)(d)}=R_{\mu\nu\rho\lambda}
e_{(a)}^{\mu}e_{(b)}^{\nu}e_{(c)}^{\rho}e_{(d)}^{\lambda}.
\end{equation}
In NP formalism, $10$ Ricci scalars ($\Phi_{ij}$ ($i,j=0,1,2$) and $\Lambda$) and $5$ complex Weyl scalars ($\Psi_0,\cdots,\Psi_4$) are introduced to describe spacetime. Four of them,
\begin{equation}
\label{Weyl-def}
\Psi_{2}=\frac{1}{6}R_{(2)(1)(2)(1)},\quad
\Psi_{3}=\frac{1}{2}R_{(2)(1)(2)(4)},\quad
\Psi_{4}=R_{(2)(4)(2)(4)},\quad\text{and}\quad
\Phi_{22}=R_{(2)(4)(2)(3)},
\end{equation}
describe all possible polarization modes of gravitational radiation. 

On the other hand, the geodesic deviation equation,
\begin{equation}
\label{geodesic-deviation-equation}
\frac{d^2\xi_{i}}{dt^2}=-R_{0i0j}\xi^{j},
\end{equation}
indicates that some components of the Riemann tensor describe the observational effects of the GWs, where $\xi^{j}$ is the deviation vector. In order to obtain the polarization modes of GW radiation, it is convenient to adopt a specific gauge for the spacetime metric. For GR theory, we know that there are just two DOFs in the GW metric through the gauge covariance, such that the transverse traceless (TT) gauge is adopted to obtain the polarization modes. However, for modified gravity, the number of polarization modes may exceed two, and the TT gauge can not be adopted arbitrarily. The NP quantities (\ref{Weyl-def}) are a set of gauge invariants. They relate to the observational effects of GWs, such that NP tetrad becomes a powerful tool to study the GW polarization modes in modified gravity. The components $R_{0i0j}$ (\ref{geodesic-deviation-equation}) can be written in terms of Ricci scalar $\Phi_{22}$ and Weyl scalars $\{\Psi_{2},\Psi_{3},\Psi_{4}\}$ (\ref{Weyl-def}) as
\begin{equation}
\label{R-component}
\begin{aligned}
R_{0101}&=\frac{1}{2}(\Phi_{22}+{\rm Re}\Psi_{4}),\quad
R_{0102}=-\frac{1}{2}{\rm Im}\Psi_{4},\quad
R_{0103}=2{\rm Re}\Psi_{3},\\
R_{0202}&=\frac{1}{2}(\Phi_{22}-{\rm Re}\Psi_{4}),\quad
R_{0203}=-2{\rm Im}\Psi_{3},\quad
R_{0303}=6\Psi_{2}.
\end{aligned}
\end{equation}
``Re" and ``Im" represent the real and imaginary parts of a quantity. On the other hand, the polarization modes are defined in terms of the components of the Riemann tensor (\ref{geodesic-deviation-equation}) in the following form,
\begin{equation}
\label{polarization-def}
R_{0i0j}=\left(\begin{array}{ccc}
\ddot{H}_{b}+\ddot{H}_{+}&
\ddot{H}_{\times} &
\ddot{H}_{x} \\
\ddot{H}_{\times} &
\ddot{H}_{b}-\ddot{H}_{+}&
\ddot{H}_{y}\\
\ddot{H}_{x}&
\ddot{H}_{y}&
\ddot{H}_{L}
\end{array}\right)
=\left(\begin{array}{ccc}
\frac{1}{2}(\Phi_{22}+{\rm Re}\Psi_{4}) &
-\frac{1}{2}{\rm Im}\Psi_{4} &
2{\rm Re}\Psi_{3} \\
-\frac{1}{2}{\rm Im}\Psi_{4} &
\frac{1}{2}(\Phi_{22}-{\rm Re}\Psi_{4})&
-2{\rm Im}\Psi_{3}\\
2{\rm Re}\Psi_{3}&-2{\rm Im}\Psi_{3}&6\Psi_{2}
\end{array}\right).
\end{equation}
In Eq.\,(\ref{polarization-def}), $H_{+}$ and $H_{\times}$ are the plus and cross modes predicted by GR. The other four independent components include two scalar modes, breathing mode $H_{b}$ and longitudinal mode $H_L$, and two vector modes, $x$-mode $H_{x}$ and $y$-mode $H_y$ \cite{Will2014test}.

\subsection{Polarization Modes}
\label{subsec:modes}
In subsection \ref{subsec:propagation-frame}, we have given the spatial components of trace-reverse metric deformation in the propagation frame, (\ref{k-propagation-frame}). Accordingly, the time component and time-spatial components are obtained by the Lorenz gauge (\ref{Lorenz}),
\begin{equation}
\label{k00-k0i-Lorenz}
\bar{k}_{0i}=\int dt(\partial_{j}\bar{k}_{ji})=-\bar{k}_{zi},
\quad\text{and}\quad
\bar{k}_{00}=\int dt(\partial_{j}\bar{k}_{j0})
=-\bar{k}_{z0}=\bar{k}_{zz}.
\end{equation}
Thus the complete metric deformation $k_{\mu\nu}$ is
\begin{equation}
\label{kmunu-Lorenz}
\begin{aligned}
&k_{\mu\nu}(\hat{\mathbf{N}})
=-\frac{4\mu x^3}{R}\delta\varpi\\
&\times\left(\begin{array}{cccc}
\frac{1}{2}\sin^2\iota\sin^2\phi
&-\frac{1}{2}\sin\iota\sin2\phi 
&-\sin\iota\cos\iota\sin^2\phi 
&-\sin^2\iota\sin^2\phi\\
-\frac{1}{2}\sin\iota\sin2\phi
&\frac{1}{2}(\cos^2\phi-\cos^2\iota\sin^2\phi)
&\frac{1}{2}\cos\iota\sin2\phi
&\frac{1}{2}\sin\iota\sin2\phi\\
-\sin\iota\cos\iota\sin^2\phi
&\frac{1}{2}\cos\iota\sin2\phi
&\frac{1}{2}(\cos^2\iota\sin^2\phi-\cos^2\phi)
& \sin\iota\cos\iota\sin^2\phi\\
-\sin^2\iota\sin^2\phi
&\frac{1}{2}\sin\iota\sin2\phi
&\sin\iota\cos\iota\sin^2\phi
&\left(1-\frac{3}{2}\cos^2\iota\right)
\sin^2\phi-\frac{1}{2}\cos^2\phi\\
\end{array}
\right).
\end{aligned}
\end{equation}
We denote the dCS part of Ricci and Weyl scalars as $\delta\Psi_{2}$, $\delta\Psi_{3}$, $\delta\Psi_{4}$, and $\delta\Phi_{22}$. The only non-zero quantity is
\begin{equation}
\label{Psi4-result}
k_+-ik_{\times}=\frac{1}{2}\int dt \int dt'\delta\Psi_{4}
=-\frac{2\mu x^3}{R}\delta\varpi
\left(\frac{1+\cos^2\iota}{2}
\cos2\phi-i\cos\iota\sin2\phi\right).
\end{equation}
Other quantities in Eq.\,(\ref{Weyl-def}) vanish, i.e., $\delta\Psi_{2}=\delta\Psi_{3}=\delta\Phi_{22}=0$. This result indicates that there is no extra polarization mode in dCS theory, which is consistent with the conclusion in Ref.\,\cite{Wagle2019}. Finally, the polarization modes of tensor radiation are
\begin{equation}
\label{H-radiation}
\begin{aligned}
H_{+,\times}(t)=h_{+,\times}(t)+k_{+,\times}(t)
&=\frac{4\mu x}{R}
\Bigg\{h_{+,\times}^{(0)}
+x^{1/2}h_{+,\times}^{(1/2)}
+x\left[h_{+,\times}^{(1)}+h_{+,\times}^{\rm(1,SO)}\right]\\
&+x^{3/2}\left[h_{+,\times}^{(3/2)}
+h_{+,\times}^{\rm(3/2,SO)}\right]
+x^{2}\left[h_{+,\times}^{(2)}
+h_{+,\times}^{\rm(2,SO)}
+h_{+,\times}^{\rm(2,SS)}
+h_{+,\times}^{\rm(2,MQ)}
+k_{+,\times}^{(2)}\right]\Bigg\}.
\end{aligned}
\end{equation}
For example, the leading-order coefficients are
\begin{equation}
\label{hpc0}
h_{+}^{(0)}=-\frac{1+\cos^2\iota}{2}\cos2\psi(t)\quad\text{and}\quad
h_{\times}^{(0)}=-\cos\iota\sin2\psi(t),
\end{equation}
and the subleading-order (0.5PN) coefficients are
\begin{equation}
\label{hpc05}
\begin{aligned}
h_{+}^{(1/2)}
&=-\frac{\sin\iota}{16}
\frac{\delta m}{m}
\left[(5+\cos^2\iota)\cos\psi(t)
-9(1+\cos^2\iota)\cos3\psi(t)\right],\\
h_{\times}^{(1/2)}
&=-\frac{3}{8}\sin\iota\cos\iota
\frac{\delta m}{m}
[\sin\psi(t)-3\sin3\psi(t)],
\end{aligned}
\end{equation}
where $\psi(t)$ is a phase related to $\phi(t)$ by
\begin{equation}
\label{psi-def}
\psi(t)=\phi(t)-2m\omega
\ln\left(\frac{\omega(t)}{\omega_{0}}\right).
\end{equation}
The constant frequency $\omega_0$ can be conveniently chosen as the entry frequency of an interferometric detector. The use of the $\psi$ instead of the actual phase $\phi$ of the source is convenient because it allows collecting the logarithmic terms that come out of the computation of the tail effects \cite{Maggiore2008}.

Additionally, the SS and MQ effects in GR are
\begin{equation}
\label{hpc-SS-MQ}
h_{+,\times}^{\rm(2,SS)}
=2\nu\left(\frac{\bm{S}_1}{m_1^2}\cdot
\frac{\bm{S}_2}{m_2^2}\right)h_{+,\times}^{(0)},
\quad\text{and}\quad
h_{+,\times}^{\rm(2,MQ)}
=\frac{1}{2}\left[\frac{m_1^2}{m^2}
\left(\frac{\bm{S}_1}{m_1^2}\right)^2
+\frac{m_2^2}{m^2}
\left(\frac{\bm{S}_2}{m_2^2}\right)^2\right]
h_{+,\times}^{(0)}.
\end{equation}
The dCS modification is
\begin{equation}
\label{kpc2}
k_{+,\times}^{(2)}=\frac{2}{3}\delta\varpi h_{+,\times}^{(0)}.
\end{equation}
Other coefficients in Eq.\,(\ref{H-radiation}) can be found in Refs.\,\cite{Blanchet2014,Kidder1995,Buonanno2013}. Notice here that we have replaced $\gamma$ and $v^2$ in Eq.\,(\ref{Hij-waveform}) with $x$ by the modified Kepler's third law (\ref{Kepler-inverse}).

From Eqs.\,(\ref{Psi4-result}) and (\ref{kpc2}), we find that there is no scalar polarization mode in the dCS gravity, although the dCS theory includes a scalar DOF. The primary reason is that the dCS theory is a quadratic theory. In the dCS action (\ref{action}), the scalar field couples with the square of the Riemann tensor. Therefore, the first-order scalar field and metric field are independent in the perturbation equation. Such that the metric deformation does not be affected by the scalar DOF, and then the breathing mode does not appear. At the same time, the Cotton tensor (\ref{Ctensor}) is traceless, such that the dCS gravity is a massless theory. Therefore, the longitudinal mode does not appear as well. Our analysis is based on the quadratic-order perturbation equation rather than PN waveforms. Such that, the above conclusion is generic for linear GWs and independent of the PN orders. Additionally, this statement is consistent with the conclusion of Ref.\,\cite{Wagle2019}, who adopted the Newman-Penrose analysis, which is independent of the PN expansion.

As we have derived in Eq.\,(\ref{kpc2}), the two polarization modes are generated symmetrically. Thus there is no parity-violating effect in the modified GW waveform. This is because we mainly consider the orbital evolution and gravitational radiation of binary systems in the wave zone and in a short time scale before merging. This time scale is very short compared with the expansion time scale of the Universe, so that the homogeneous background scalar field in the dCS theory does not need to be considered. The amplitude birefringence, a kind of important parity-violating effect, will be investigated in the propagation zone (see Section \ref{sec:propagation}). 

\section{Energy Flux and Orbital Evolution}
\label{sec:flux}
% This section will give the energy flux carried by scalar and tensor radiation, which are shown in subsection \ref{subsec:scalarflux} and \ref{subsec:tensorflux}, respectively.
\subsection{Scalar Field}
\label{subsec:scalarflux}
From the EMT of the scalar field (\ref{EMTtheta}), the time-spatial component is \cite{Yagi2012pn}
\begin{equation}
\label{T0i-theta}
T_{0i}^{(\vartheta)}
=\beta\dot{\vartheta}(\partial_{i}\vartheta),
\end{equation}
and then the scalar energy flux is
\begin{equation}
\label{scalar-flux}
\mathcal{F}_{S}=-\oint_{\partial\Omega}\left\langle
\hat{N}_{i}T_{0i}^{(\vartheta)}\right\rangle R^2 d\Omega
=\beta R^2\oint_{\partial\Omega}\langle\dot{\vartheta}^2\rangle d\Omega,
\end{equation}
where $\left\langle\cdots\right\rangle$ represents averaging over a period. Then, substituting the scalar waveform (\ref{theta-radiation}) into Eq.\,(\ref{scalar-flux}), using the integral formula,
\begin{equation}
\label{NNNN-integration}
\int \hat{N}_{i}\hat{N}_{j}\hat{N}_{k}\hat{N}_{m}
d\Omega
=\frac{4\pi}{15}\left(\delta_{ij}\delta_{km}+\delta_{ik}\delta_{jm}+\delta_{im}\delta_{jk}\right),
\end{equation}
and averaging formula, $\left\langle v^2\right\rangle=\left\langle(\hat{\bm{v}}\cdot\bm{\Delta})^2\right\rangle=0$, we get the final scalar energy flux,
\begin{equation}
\label{scalar-flux-result}
\mathcal{F}_{S}=\frac{32}{5}\nu^2x^7\left(
\frac{25}{24576}\frac{\zeta}{\nu^2}\bm{\Delta}^2\right).
\end{equation}
The symbol $\bm{\Delta}$ has been defined in Eq.\,(\ref{theta-radiation}). This result (\ref{scalar-flux-result}) is consistent with that given by Ref.\,\cite{Yagi2016e}, while the term $27\left\langle(\bm{\Delta}\cdot\bm{v})^2\right\rangle$ is omitted because of spin-aligned assumption.

\subsection{Tensor Field}
\label{subsec:tensorflux}
In terms of metric tensor (\ref{metric-g}), the time-spatial component of the EMT of the gravitational field is
\begin{equation}
\label{T0i-H}
T_{0i}^{(H)}
=\frac{1}{32\pi}\left\langle
(\partial_{0}\bar{H}^{\rm TT}_{jk})
(\partial_{i}\bar{H}^{\rm TT}_{jk})\right\rangle.
\end{equation}
The radiation energy flux is written as
\begin{equation}
\label{tensor-flux}
\mathcal{F}_{H}=-\oint_{\partial\Omega}\left\langle
\hat{N}_{i}T_{0i}^{(H)}\right\rangle R^2 d\Omega
\approx\frac{R^2}{32\pi}\oint_{\partial\Omega}
\left[\langle\dot{h}^{\rm TT}_{jk}\dot{h}^{\rm TT}_{jk}\rangle
+2\langle\dot{h}^{\rm TT}_{jk}\dot{k}^{\rm TT}_{jk}\rangle\right]d\Omega.
\end{equation}
The first term in the square bracket is the GR part, while the second represents the contributions from the dCS theory. We rewrite them as
\begin{equation}
\label{tensor-flux-GR-dCS}
\bar{\mathcal{F}}_{H}
=\frac{1}{16\pi}R^2\oint_{\partial\Omega}
\left[\langle\dot{h}_{+}^2\rangle
+\langle\dot{h}_{\times}^2\rangle\right]d\Omega\quad\text{and}\quad
\delta\mathcal{F}_{H}
=\frac{1}{8\pi}R^2\oint_{\partial\Omega}
\left[\langle\dot{h}_{+}\dot{k}_{+}\rangle
+\langle\dot{h}_{\times}\dot{k}_{\times}\rangle\right]d\Omega
\end{equation}
in terms of polarization modes (\ref{Hij-waveform}). The GR part has been fully calculated in previous work \cite{Kidder1995,Blanchet2002,Blanchet2006,Blanchet2014,Bohe2015,Cho2021,Cho2022}. Finally, we obtain the dCS modification of the energy flux carried by tensor radiation,
\begin{equation}
\label{tensor-flux-dCS-result}
\delta\mathcal{F}_{H}
=\frac{32}{5}\nu^2x^7\left(\frac{4}{3}\delta\varpi\right).
\end{equation}
We have used $\langle\sin^2[2\phi(t)]\rangle=\langle\cos^2[2\phi(t)]\rangle=1/2$. Combined Eqs.\,(\ref{scalar-flux-result}) and (\ref{tensor-flux-dCS-result}), the total energy flux is written as
\begin{equation}
\label{flux}
\mathcal{F}
=\frac{32}{5}\nu^2x^5
\left[1+x\bar{\mathscr{F}}_{1}
+x^{3/2}\bar{\mathscr{F}}_{1.5}
+x^2(\bar{\mathscr{F}}_{2}+\delta\mathscr{F})
\right].
\end{equation}
The 1PN, 1.5PN, and 2PN coefficients are given by
\begin{equation}
\label{mathcal-F1}
\bar{\mathscr{F}}_{1}=-\frac{1247}{336}
-\frac{35}{12}\nu,
\end{equation}
\begin{equation}
\label{mathcal-F15}
\bar{\mathscr{F}}_{1.5}=4\pi
-\left(4s+\frac{5}{4}\delta\sigma\right),
\end{equation}
and
\begin{equation}
\label{mathcal-F2}
\mathscr{F}_{2}=-\frac{44711}{9072}
+\frac{9271}{504}\nu
+\frac{65}{18}\nu^2
+\frac{31}{8}\nu
\left(\frac{\bm{S}_1}{m_1^2}
\cdot\frac{\bm{S}_{2}}{m_2^2}\right)
+\frac{33}{16}
\left[\frac{m_1^2}{m^2}
\left(\frac{\bm{S}_{1}}{m_1^2}\right)^2
+\frac{m_2^2}{m^2}
\left(\frac{\bm{S}_{2}}{m_2^2}\right)^2\right],
\end{equation}
respectively. The factor $4\pi$ in (\ref{mathcal-F15}) comes from tail terms of GW radiation. The dCS modification of energy flux is
\begin{equation}
\label{mathcal-delta-F}
\begin{aligned}
\delta\mathscr{F}=
\frac{25}{24576}\frac{\zeta}{\nu^2}\bm{\Delta}^2
+\frac{4}{3}\delta\varpi
=\zeta
\left\{\frac{4775}{12288}\frac{1}{\nu}
\left(\frac{\bm{S}_{1}}{m_1^2}
\cdot\frac{\bm{S}_{2}}{m_2^2}\right)
-\frac{38417}{172032}\left[\frac{m^2}{m_1^2}
\left(\frac{\bm{S}_{1}}{m_1^2}\right)^2
+\frac{m^2}{m_2^2}
\left(\frac{\bm{S}_{2}}{m_2^2}\right)^2\right]\right\}.
\end{aligned}
\end{equation}
Previous work has also given the dCS flux. However, the scalar flux given by Ref.\,\cite{Yagi2012pn} is not complete, and the metric deformation flux is absent in Ref.\,\cite{Yagi2016e} because of unreasonable magnitude estimation. Similar to the discussion after Eq.\,(\ref{delta-epsilon}), for equal-mass and equal spin systems, the ratio of radiation energy flux caused by SS and MQ effect is about $0.14$. That is, the correction of the MQ effect on radiant energy flux and orbital evolution is much more obvious than that of the SS effect.

\subsection{Orbital Evolution}
\label{subsec:orbitevolution}
The energy flux carried by radiation comes from the binding energy of the binary system. Using the conservation of energy, $\mathcal{F}=-\dot{E}$, the differential equation of frequency evolution with time can be established in terms of the dimensionless frequency $x$, 
\begin{equation}
\label{x-dot}
\dot{x}=\frac{64}{5}\frac{\nu}{m}x^5
\left[1+x\bar{\Omega}_{1}
+ x^{3/2}\bar{\Omega}_{1.5}
+x^2\left(\bar{\Omega}_{2}
+\delta\Omega\right)\right].
\end{equation}
The PN coefficients are
\begin{equation}
\label{Omega-1}
\bar{\Omega}_{1}
=-\frac{743}{336}-\frac{11}{4}\nu,
\end{equation}
\begin{equation}
\label{Omega-15}
\bar{\Omega}_{1.5}
=4\pi-\left(\frac{47}{3}s
+\frac{25}{4}\delta\sigma\right),
\end{equation}
\begin{equation}
\label{Omega-2}
\bar{\Omega}_{2}
=\frac{34103}{18144}
+\frac{13661}{2016}\nu
+\frac{59}{18}\nu^2
+\frac{79}{8}\nu
\left(\frac{\bm{S}_1}{m_1^2}
\cdot\frac{\bm{S}_{2}}{m_2^2}\right)
+\frac{81}{16}
\left[\frac{m_1^2}{m^2}
\left(\frac{\bm{S}_{1}}{m_1^2}\right)^2
+\frac{m_2^2}{m^2}
\left(\frac{\bm{S}_{2}}{m_2^2}\right)^2\right].
\end{equation}
The dCS modification takes the form of
\begin{equation}
\label{delta-Omega}
\delta{\Omega}=\zeta
\left\{\frac{11975}{12288}\frac{1}{\nu}
\left(\frac{\bm{S}_1}{m_1^2}
\cdot\frac{\bm{S}_2}{m_2^2}\right)
-\frac{96305}{172032}
\left[\frac{m^2}{m_1^2}
\left(\frac{\bm{S}_1}{m_1^2}\right)^2
+\frac{m^2}{m_2^2}
\left(\frac{\bm{S}_2}{m_2^2}\right)^2\right]\right\}.
\end{equation}
With the energy dissipation, the binding energy of the binary system decreases. Combined with modified Kepler's third law, one can calculate the ppE parameters directly without using the waveform. The orbital radius of the binary becomes smaller and smaller, and thus the angular frequency, ($\omega$ or $x$), becomes larger and larger. When the orbital radius is close to the gravitational radius of the BH, the near-zone gravitational field is so strong that the PN approximation is no longer valid. Let the merging time of the binary BHs be $t_c$, and define the time from merging time as $\tau=t_c-t$. In terms of $\tau$, above differential equation (\ref{x-dot}) is solved perturbatively as
\begin{equation}
\label{tau-solution}
\tau=\int_{\tau}^{0}d\tau
=\frac{5}{256}\frac{m}{\nu}\frac{1}{x^4}\\
\left[1-\frac{4}{3}x\bar{\Omega}_1
-\frac{8}{5}x^{3/2}\bar{\Omega}_{1.5}
+2x^2\left(\bar{\Omega}_1^2-\bar{\Omega}_2
-\delta\Omega\right)\right].
\end{equation}
Inversely, the solution to $x(\tau)$ is
\begin{equation}
\label{x-solution}
x=\frac{1}{4}\Theta^{-1/4}\left\{1
-\frac{1}{12}\bar{\Omega}_{1}\Theta^{-1/4}
-\frac{1}{20}\bar{\Omega}_{1.5}\Theta^{-3/8}
+\frac{1}{288}
\left[8\bar{\Omega}_{1}^2-9(\bar{\Omega}_{2}
+\delta\Omega)\right]\Theta^{-1/2}\right\},
\end{equation}
where the dimensionless time is defined as $\Theta(\tau)\equiv\nu\tau/5m$, relating to the orbital phase by
\begin{equation}
\label{x-phi}
x^{3/2}=-\frac{\nu}{5}\frac{d\phi}{d\Theta}.
\end{equation}
Integrating Eq.\,(\ref{x-phi}) gives
\begin{equation}
\label{phi-Theta}
\begin{aligned}
\phi(\Theta)=\phi_{c}
-\frac{5}{\nu}\int^{\Theta}_{0}
x^{3/2}(\Theta)d\Theta
=\phi_{c}-\frac{1}{\nu}\Theta^{5/8}
\left[1+\bar{\kappa}_{1}\Theta^{-1/4}
+\bar{\kappa}_{1.5}\Theta^{-3/8}
+(\bar{\kappa}_{2}+\delta\kappa)
\Theta^{-1/2}\right],
\end{aligned}
\end{equation}
where $\phi_c$ is the initial orbital phase. The higher-order PN coefficients are given by
\begin{equation}
\label{kappa-1}
\bar{\kappa}_{1}=\frac{3715}{8064}+\frac{55}{96}\nu,
\end{equation}
\begin{equation}
\label{kappa-15}
\bar{\kappa}_{1.5}=-\frac{3\pi}{4}
+\left(\frac{47}{16}s
+\frac{75}{64}\delta\sigma\right),
\end{equation}
and
\begin{equation}
\label{kappa-2}
\bar{\kappa}_{2}=\frac{9275495}{14450688}
+\frac{284875}{258048}\nu
+\frac{1855}{2048}\nu^2
-\frac{1185}{512}\nu
\left(\frac{\bm{S}_1}{m_1^2}
\cdot\frac{\bm{S}_2}{m_2^2}\right)
-\frac{1215}{1024}
\left[\frac{m_1^2}{m^2}
\left(\frac{\bm{S}_{1}}{m_1^2}\right)^2
+\frac{m_2^2}{m^2}
\left(\frac{\bm{S}_{2}}{m_2^2}\right)^2\right],
\end{equation}
The dCS correction is
\begin{equation}
\label{delta-kappa}
\delta\kappa=
\zeta\left\{-\frac{59875}{262144}\frac{1}{\nu}
\left(\frac{\bm{S}_1}{m_1^2}
\cdot\frac{\bm{S}_2}{m_2^2}\right)
+\frac{481525}{3670016}\left[\frac{m^2}{m_1^2}
\left(\frac{\bm{S}_1}{m_1^2}\right)^2
+\frac{m^2}{m_2^2}
\left(\frac{\bm{S}_2}{m_2^2}\right)^2\right]\right\}.
\end{equation}
Similar to the discussions around Eqs.\,(\ref{delta-epsilon}) and (\ref{mathcal-delta-F}), for the equal-mass and equal spin binary BH system, the ratio of correction of orbital phase caused by SS and MQ effect is about $0.87$. The contribution from the SS effect is comparable with the MQ effect for orbital phase evolution.

\section{Frequency-Domain Waveform and ppE parameters}
\label{sec:FDwaveform}
In extracting GWs from noisy data, it is customary to employ the frequency-domain gravitational waveform in Fourier space, which can be computed analytically via the stationary phase approximation \cite{Maggiore2008,Arun2009,XingZhang2017SMG,XingZhang2019SMG2,TanLiu2020BD}. For preparation, let us first write the detected GW signal in the following form, 
\begin{equation}
\label{Ht}
\begin{aligned}
H(t)=H_{+}F_{+}+H_{\times}F_{\times}
=\frac{4\mu}{R}x(t_r)\sum_{n=1}^{6}\sum_{k=0}^{4}
x^{k/2}(t_r)
\times\left\{c_{n,k}\cos\left[n\phi(t)\right]
+s_{n,k}\sin\left[n\phi(t)\right]\right\},
\end{aligned}
\end{equation}
where $F_{+}$ and $F_{\times}$ are antenna pattern functions of GW detectors \cite{Maggiore2008}, the coefficients $c_{n,k}$ and $s_{n,k}$ are a set of constants depending on the mass and spin of binary BHs. The index $k$ denotes the order or PN expansion, and $n$ is the multiple of GW frequency relative to the orbital frequency $\omega$. These constants can be read out from the gravitational waveforms \cite{Blanchet1995,Kidder1995,Buonanno2013,Bohe2015}. The Fourier transformation of the signal (\ref{Ht}) is 
\begin{equation}
\label{Hf}
\begin{aligned}
\tilde{H}(f)=\int_{-\infty}^{\infty}
H(t)e^{i2\pi ft}dt
=\frac{2\mu}{R}e^{i2\pi fR}\sum_{n=1}^{6}\sum_{k=0}^{4}
\int_{-\infty}^{\infty}dtx^{k/2+1}e^{i2\pi f t}
\left[(c_{n,k}-is_{n,k})e^{in\phi}
+(c_{n,k}+is_{n,k})
e^{-in\phi}\right],
\end{aligned}
\end{equation}
where $f$ is the oscillation frequency of GW signals. Eq.\,(\ref{Hf}) is an oscillation integration, mainly contributed by the value of the integrated function at the stationary point. The stationary points are determined by equations $n\dot{\phi}=2\pi f$ and $n\dot{\phi}=-2\pi f$. The former equation gives
\begin{equation}
\label{Theta-n-stationary}
\begin{aligned}
\Theta^{*}_n=\frac{1}{256}u_n^{-8/3}\left\{
1+\frac{32}{5}\kappa_{1}u_n^{2/3}
+\frac{128}{15}\kappa_{1.5}u_n
+\frac{64}{75}\left[3\kappa_{1}^2
+10(\kappa_2+\delta\kappa)\right]u_n^{4/3}\right\},
\end{aligned}
\end{equation}
while the latter will not give a stationary point. Eq.\,(\ref{Theta-n-stationary}) introduces another dimensionless frequency as
\begin{equation}
\label{un-def}
u_{n}\equiv\frac{2}{n}\pi mf
\end{equation}
for convenience. The frequency-domain waveform is thus obtained, 
\begin{equation}
\begin{aligned}
\tilde{H}(f)&=\sqrt{\frac{5}{12}}\frac{m^2}{R}
\sum_{n=1}^{6}\sum_{k=0}^{4}\sqrt{\frac{\pi\nu}{n}}
(c_{n,k}+is_{n,k})
\times u_n^{(2k-7)/6}e^{i\Psi_n(f)}\\
&\qquad\times\left\{1+\frac{12}{5}\kappa_{1}u_n^{2/3}
+\frac{8}{3}\kappa_{1.5}u_n
+\frac{32}{15}\left[(\kappa_2+\delta\kappa)
-\frac{21}{20}\kappa_{1}^2\right]u_n^{4/3}\right\}.
\end{aligned}
\end{equation}
The frequency-domain phase is
\begin{equation}
\Psi_n(f)
=2\pi f(t_c+R)-\frac{\pi}{4}-n\phi_{c}
+\frac{3}{256}\frac{n}{\nu}u_{n}^{-5/3}\left\{
1+\frac{32}{3}\kappa_{1}u_{n}^{2/3}
+\frac{64}{3}\kappa_{1.5}u_{n}
+\frac{64}{15}\left[3\kappa_{1}^2
+10(\kappa_2+\delta\kappa)\right]u_{n}^{4/3}\right\}.
\end{equation}
So far, we have completed the calculation of ready-to-use gravitational waveform radiated from binary slowly-rotating BHs in dCS gravity.

We rewrite this frequency-domain waveform in terms of ppE parameters,
\begin{equation}
\label{Hf-ppE}
\tilde{H}(f)=\tilde{h}(f)\left[1
+\alpha_{\rm g}u^{a_{\rm g}}\right]
\exp\left[i\beta_{\rm g}u^{b_{\rm g}}\right],
\end{equation}
where $\tilde{h}(f)$ is the GR waveform. $u\equiv(\pi\mathcal{M}f)^{1/3}$ is another dimensionless frequency, with chirp mass defined as $\mathcal{M}=\nu^{3/5}m$. The ppE parameters are
\begin{equation}
a_{\rm g}=4,\quad b_{\rm g}=-1,
\end{equation}
and
\begin{equation}
\label{ppE-alpha-beta-g}
\begin{aligned}
\alpha_{\rm g}
&=\zeta\nu^{-4/5}\left\{
-\frac{7175}{24576}\frac{1}{\nu}
\left(\frac{\bm{S}_1}{m_1^2}
\cdot\frac{\bm{S}_2}{m_2^2}\right)
+\frac{57713}{344064}
\left[\frac{m^2}{m_1^2}
\left(\frac{\bm{S}_1}{m_1^2}\right)^2
+\frac{m^2}{m_2^2}
\left(\frac{\bm{S}_2}{m_2^2}\right)^2\right]\right\},\\
\beta_{\rm g}
&=\zeta\nu^{-4/5}\left\{
-\frac{59875}{262144}\frac{1}{\nu}
\left(\frac{\bm{S}_1}{m_1^2}
\cdot\frac{\bm{S}_2}{m_2^2}\right)
+\frac{481525}{3670016}
\left[\frac{m^2}{m_1^2}
\left(\frac{\bm{S}_1}{m_1^2}\right)^2
+\frac{m^2}{m_2^2}
\left(\frac{\bm{S}_2}{m_2^2}\right)^2\right]\right\}.
\end{aligned}
\end{equation}
The dCS theory not only corrects the phase of GWs predicted by GR but also corrects its amplitude. This result (\ref{Hf-ppE}) gives the gravitational waveform radiated by the spin-aligned circular-orbits binary BHs in the wave zone, where the effects of cosmological background evolution and parity-violating amplitude birefringence effect are not considered. This waveform can be directly used to extract GW signals, estimate the constrain ability of the future GW detector on dCS theory and constrain the dCS theory through the current observation data. The same results will be given by a simple formulas, $\alpha_{\rm g}=(2/3)\delta\varpi-(1/2)\delta\Omega$ and $\beta_{\rm g}=(15/64)\delta\Omega$, provided by Ref.\,\cite{Tahura2018}, in which the consistent result is given.

\section{Propagation Effect of GWs}
\label{sec:propagation}
As mentioned earlier, the inspiraling binary BH system generates only two polarization modes of GWs. Due to the time scale of the binary merge being too short, the parity-violating effect will not appear in the wave zone. However, when the GW enters the propagation zone, the background scalar field will induce the amplitude birefringence effect, and the two polarization modes will transform each other.
% Subsection \ref{subsec:circular-polarization-equation} reviews the evolution equation of GW circular polarization modes. The amplitude birefringence is derived in subsection \ref{subsec:amplitude-birefringence}. Finally, the detected frequency-domain wavefrom is given in subsection \ref{subsec:detected-signal}.

\subsection{Evolution Equation of Circular Polarization Modes}
\label{subsec:circular-polarization-equation}
The background metric is assumed as flat Robertson-Walker geometry, 
\begin{equation}
ds^2=g_{\mu\nu}^{\rm(B)}dx^{\mu}dx^{\nu}
=a^2(\eta)(-d\eta^2+d\mathbf{x}^2).
\end{equation}
$\eta$ denotes the conformal time, and $a(\eta)$ denotes the scale factor of the Universe. Throughout this section, we set the present scale factor $a_{0}\equiv a(\eta_{0})=1$. conformal time relates to the cosmic time $t$ by $dt=a(\eta)d\eta$. The metric perturbation is also denoted by $H_{\mu\nu}$, which satisfies the TT gauge
\begin{equation}
H_{0\mu}=0,\quad
\partial^{i}H_{ij}=0,\quad
H=g^{ij}H_{ij}=0.
\end{equation}
From the total metric $g^{\rm(B)}_{\mu\nu}+H_{\mu\nu}$ and field equation (\ref{tensor-equation}), the evolution equation of $H_{\mu\nu}$ is given by
\begin{equation}
\label{tensor-equation-RW}
H''_{jk}+2\mathcal{H}H'_{jk}-(\Box^2_{\eta}H_{jk})
+\frac{16\pi\alpha}{a^2}\epsilon^{sm}_{\ \ \ (j}
\partial_{m}\left[\vartheta'_0H_{sk)}''
+\vartheta''_0H_{sk)}'
-\vartheta'_0(\Box^2_{\eta}H_{sk)})\right]=0.
\end{equation}
In above Eq.\,(\ref{tensor-equation-RW}), $\mathcal{H}\equiv a'/a$ is the Hubble parameter. A prime $(\prime)$ denotes the derivative with respect to the conformal time $\eta$. $\vartheta_{0}$ is the background scalar field, which homogeneously distributes in the entire Universe. It is just a function of cosmic time and changes with the expansion of the Universe. Because of the background scalar field, GWs exhibit an amplitude birefringence effect during propagation, which violates the prediction by GR. We decompose the metric perturbation according to different wave vectors and divide it into left- and right-hand polarization modes,
\begin{equation}
\label{circular-polarization-decomposition}
H_{jk}=\sum_{A=L,R}
\int\frac{d^3\bm{\kappa}}{(2\pi)^3}
H_{A}(\eta,\kappa_{i})
\exp\left(i\bm{\kappa}\cdot\mathbf{x}\right)
e_{jk}^{A}.
\end{equation}
The subscripts $A=L,R$ represent two kinds of circular modes. $\bm{\kappa}$ is the wavevector of propagating GWs, and we define $\kappa^2\equiv\bm{\kappa}\cdot\bm{\kappa}$ and $\hat{\bm{\kappa}}\equiv\bm{\kappa}/\kappa$. The circular polarization basis tensors are defined as $e_{ij}^{L,R}=(1/\sqrt{2})(e_{ij}^{+}\mp ie_{ij}^{\times})$. $e^{+,\times}_{ij}$ describe the plus and cross polarization modes. Additionally, the basis tensors satisfy
\begin{equation}
\hat{\kappa}_{m}\epsilon^{sm}_{\ \ \ j}e^{A}_{sk}
=i\rho_{A}e_{jk}^{A}
\end{equation}
with $\rho_{L}=-1$ and $\rho_{R}=+1$.
Decomposition (\ref{circular-polarization-decomposition}) gives two independent equations of circular polarizations,
\begin{equation}
\label{left-right-polarization-equation}
H''_{A}
+2\mathcal{H}H'_{A}
+\kappa^2H_{A}
-\frac{16\pi\alpha\kappa}{a^2}
\rho_{A}\left[\vartheta'_0H_{A}''
+\vartheta''_0H_{A}'
+\kappa^2\vartheta'_0H_{A}\right]=0,
\end{equation}
or equivalently,
\begin{equation}
\label{left-right-polarization-equation-nuA}
H_{A}''+\left(2+\nu_{A}\right)\mathcal{H}H_{A}'
+\kappa^2H_{A}=0,
\quad\text{with}\quad
\nu_{A}=\frac{(16\pi\alpha)\rho_{A}\kappa
(2\mathcal{H}\vartheta'_0-\vartheta''_0)/(a^2\mathcal{H})}
{1-(16\pi\alpha)\rho_{A}\kappa\vartheta'_0/a^2}.
\end{equation}
$\nu_{A}$ depends on the modes and describes the amplitude birefringence effect, that is, the amplitudes of left- and right-hand waves will gradually increase (decrease) during propagation. It can be seen that the dCS theory does not modify the wave number of GWs, so GWs still travel in vacuum at the speed of light, whether left- or right-hand polarization mode.

\subsection{Amplitude Birefringence}
\label{subsec:amplitude-birefringence}
The amplitude corrections to the gravitational waveform arising from the parameters $\nu_{A}$ are given from the evolution equation (\ref{left-right-polarization-equation-nuA}). We further decompose circular polarizations as
\begin{equation}
\label{HA-amplitude-correction}
H_{A}=\mathcal{A}_{A}
e^{-i\Phi_{A}(\eta)}e^{-i\theta_{A}(\eta)}.
\end{equation}
$\mathcal{A}_{A}$ and $\Phi_{A}$ represent the original amplitude and phase of circular polarization waveforms. The phase factor $e^{-i\theta}$ denotes amplitude correction induced by parameter $\nu_{A}$. Substituting Eq.\,(\ref{HA-amplitude-correction}) into the perturbation equation, we give the equation of $\theta_{A}$,
\begin{equation}
\label{theta-equation}
i\theta''+2\theta'\Phi'
+\theta^{\prime2}
+i\left(2+\nu_{A}\right)\mathcal{H}\theta'
+i\mathcal{H}\nu_{A}\Phi'=0,
\end{equation}
Because of $\theta_{A}\sim\nu_{A}$ and $d/d\eta\sim\mathcal{H}$, we have $\theta'_{A}\sim\mathcal{H}\theta$ and $\theta''_{A}\sim\mathcal{H}^2\theta$. At the same time, the wavelength is assumed to be much smaller than the scale of the Universe, such that $\kappa\gg\mathcal{H}$. Finally, Eq.\,(\ref{theta-equation}) is approximated as 
\begin{equation}
2\theta'\Phi'+i\mathcal{H}\nu_{A}\Phi'=0.
\end{equation}
This is integrated from the conformal time of the GW source to the present time. The solution to $\theta_{A}$ is obtained,
\begin{equation}
\theta_{A}=-\frac{i}{2}\int_{\eta_{s}}^{\eta_0}
\mathcal{H}\nu_{A}d\eta.
\end{equation}
Compared with the original one, $\mathcal{A}_{A}e^{-i\Phi_{A}(\eta)}$, the detected waveform is written as
\begin{equation}
H_{A}^{\rm(o)}=H_A(1+\delta H_{A}),
\end{equation}
where $\delta H_{A}$ is the amplitude correction for each polarization mode, which is calculated as follows,
\begin{equation}
1+\delta H_{A}=e^{-i\theta_{A}}
=1-\frac{1}{2}\int_{\eta_{s}}^{\eta_0}
\mathcal{H}\nu_{A}d\eta
=1-\ln
\sqrt{\frac{1-(16\pi\alpha)\rho_{A}
\kappa[\vartheta'_0(\eta_{0})/a^2(\eta_{0})]}
{1-(16\pi\alpha)\rho_{A}
\kappa[\vartheta'_0(\eta_{s})/a^2(\eta_{s})]}}
\approx1+\rho_{A}\xi u^3,
\end{equation}
where
\begin{equation}
\xi\equiv\frac{16\pi\alpha}{\mathcal{M}}
\left[\vartheta'_0(\eta_{0})
-(1+z)^2\vartheta'_0(\eta_{s})\right].
\end{equation}
The cosmological redshift of the wave source is defined as $z\equiv1/a(\eta_{s})-1$. In terms of plus and cross modes, the detected polarizations are written as
\begin{equation}
\label{Hpco}
H_{+}^{\rm(o)}=H_+-i\xi u^3H_{\times},\quad
H_{\times}^{\rm(o)}=H_{\times}+i\xi u^3H_+.
\end{equation}
Amplitude birefringence will enter at least 1.5PN order correction of the waveform. Like the gravitational Faraday rotation of GWs \cite{GFR}, the polarization tensor will rotate an angle on the polarization plane, and the plus and cross modes then transform each other.

\subsection{The Detected Signal}
\label{subsec:detected-signal}
The detected GW signal is a linear combination of plus and cross modes, 
\begin{equation}
\label{Ho}
H^{\rm(o)}=H_{+}^{\rm(o)}F_{+}+H_{\times}^{\rm(o)}F_{\times}=H_+F_{+}+H_{\times}F_{\times}
-i\xi u^{3}\left(H_{\times}F_{+}
-H_{+}F_{\times}\right).
\end{equation}
During the process of GW generation, the dCS theory influences the waveform at 2PN order. However, in the propagation zone, the leading-order modification by amplitude birefringence enters the 1.5PN waveform, while the modified 0.5PN waveform enters the 2PN order. In Fourier space, the waveform is modified as
\begin{equation}
\label{Hof}
\tilde{H}^{\rm(o)}(f)=\tilde{H}(f)
-i\xi\tilde{h}_{\xi}(f).
\end{equation}
The first term is given in Eq.\,(\ref{Hf}). Up to 2PN order, the second term is
\begin{equation}
\tilde{h}_{\xi}(f)
=\sqrt{\frac{5\pi}{12}}
\frac{m^2}{R}
\nu^{6/5}u^{-1/2}
\Bigg\{\alpha_{02}
e^{i\beta_2(f)}
+2^{-5/6}\nu^{-1/5}u
\left[\alpha_{11}
e^{i\beta_1(f)}
+\alpha_{13}
e^{i\beta_3(f)}\right]\Bigg\}e^{2\pi f(t_c+R)-\pi/4},
\end{equation}
without dCS modification. The amplitude correction parameters are
\begin{subequations}
\begin{align}
\alpha_{02}
&=\frac{1}{\sqrt{2}}
\left[\frac{1}{2}(1+\cos^2\iota)F_{\times}
-i\cos\iota F_{+}\right],\\
\alpha_{11}&=\frac{1}{4}
\frac{\delta m}{m}\sin\iota
\left[(5+\cos^2\iota)F_{\times}
-3i\cos\iota F_{+}\right],\\
\alpha_{13}
&=-\frac{9}{4}\sqrt[3]{3}
\frac{\delta m}{m}\sin\iota
\left[(1+\cos^2\iota)F_{\times}
-i\cos\iota F_{+}\right].
\end{align}
\end{subequations}
The phase correction parameter is
\begin{equation}
\beta_n(f)=-n\left[\Phi_{c}
+\frac{3}{256}(n/2)^{5/3}
u^{-5}\right].
\end{equation}

The dCS modification during GW generation and propagation can be put together. The detected frequency-domain signal is then written as
\begin{equation}
\label{Hof-hof}
\tilde{H}^{\rm(o)}(f)=\tilde{h}(f)
(1+\alpha_{\rm g} u^{4})
\exp(i\beta_{\rm g} u^{-1})
-i\xi\tilde{h}_{\xi}(f).
\end{equation}
The parameters $\alpha_{\rm g}$ and $\beta_{\rm g}$ describe correction to the generation process, which enters 2PN approximation, measured by coupling $\alpha$ or $\zeta$. Ref.\,\cite{Silva2021} constrains $\sqrt{\alpha}\leq8.5\,{\rm km}$ by multi-messenger neutron star observations. Another term in Eq.\,(\ref{Hof-hof}) describes amplitude correction induced by parity-violating effect, which enters 1.5PN and 2PN corrections. This effect is measured by the coupling parameter and the time derivative of the background scalar field, $\alpha\vartheta'_{0}$. Ref.\,\cite{Smith2008} used solar-system measurements of frame-dragging from LAGEOS and Gravity Probe B to bound $\ell_{0}\equiv\alpha\dot{\vartheta}_{0}/16\pi\leq 3000\,{\rm km}$. Ref.\,\cite{Maria2022} improved this constraint to $\ell_{0}\leq 1000\,{\rm km}$ using GWTC-2 catalog \cite{GWTC2}.

\section{Conclusion and Discussion}
\label{sec:conslusion}
DCS theory is a parity-violating gravitational theory that has recently attracted more and more attention. %The polarization modes of GWs are studied in the generation and propagation process in dCS gravity. 
In this article, we calculate the gravitational waveform radiated by binary slowly-rotating BHs,
where the polarization modes of GWs are studied in the generation and propagation process. The EOM is established from the modified MPD equation, which is a powerful tool to describe the motion of spinning point mass based on the EFT method. Compared with Ref.\,\cite{Yagi2012gw}, we emphasize that the EOM is derived for spin-aligned quasi-circular orbits case, in which the SAM of each object, $\bm{S}_{A}$, is conserved, without spin precession. The complete relative acceleration of the binary system, $\bm{a}$, is given in Eq.\,(\ref{a-spin-aligned}), including non-spinning effect, SO, SS, and MQ coupling. The dCS modification, $\delta\bm{a}$, is similar to SS and MQ effects predicted by GR, which enters 2PN corrections.

Based on the EOM of binary, the scalar and tensor-deformation radiations, $\vartheta$ and $\bar{k}_{ij}$, are calculated through the multipole moment formula. Through NP tetrad, the polarization modes are investigated in this work. We find that DCS theory cannot produce any extra scalar and vector modes since dCS is a quadratic and massless theory, although a scalar DOF is introduced in the action. The plus and cross modes, $k_{+}$ and $k_{\times}$, are presented in Eq.\,(\ref{Psi4-result}). In the wave zone, there is no parity-violating effect for plus and cross modes because of the short time scale of the binary merger, in which the expansion of the Universe and homogeneous background scalar field are not taken into consideration.

The energy flux carried by radiation consists of scalar and tensor fluxes. Compared with Refs.\,\cite{Yagi2012gw,Yagi2016e}, the flux of metric deformation, $\delta\mathcal{F}_{H}$, is also included, whose leading-order modification is at 2PN order, same as the scalar flux, rather than 3PN. Following the standard PN method, the orbital decay $\dot{x}$, frequency-domain waveform $\tilde{H}(f)$ and improved ppE parameters $(\alpha_{\rm p},\beta_{\rm p})$ are also given in Eqs.\,(\ref{x-dot}), (\ref{Hf}) and (\ref{Hf-ppE}). The GWs need to travel for a long time before reaching the detector. The effects of cosmological metric and background scalar field $\vartheta_{0}$ are significant in this process. In dCS gravity, the amplitudes of the left-hand circular polarization mode of GW, $H_{L}$, will increase (decrease) during its propagation, while the right-hand mode, $H_{R}$, will decrease (increase), which is called amplitude birefringence (\ref{Hpco}). Thus, more amplitude and phase corrections are introduced into the detected frequency-domain waveform (\ref{Hof-hof}). This effect enters 1.5PN correction, which is more obvious than that produced in the generation process. This result may be helpful for the joint constraint on the background scalar field $\theta_{0}$ and dCS coupling parameter $\alpha$. The future third-generation ground-based detectors, including Einstein Telescope and Cosmic Explorer, and the space-based GW detectors, like LISA, TianQin and Taiji, will hopefully detect the higher PN waveform \cite{ChangfuShi2022}. Our results can be used for GW signal extraction, detector ability prediction, and gravitational test in future work.

\begin{acknowledgments}
We would like to thank Kent Yagi, Alessandra Buonanno, Guillaume Faye, Nicholas Loutrel, Sharaban Tahura, Shaoqi Hou, Rui Niu and Yupeng Zhang for helpful discussions and comments. This work is supported by the National Key R\&D Program of China Grant No.\,2022YFC2200100 and 2021YFC2203102, NSFC No.\,12273035 and 11903030 the Fundamental Research Funds for the Central Universities under Grant No.\,WK2030000036 and WK3440000004, and the science research grants from the China Manned Space Project with No.CMS-CSST-2021-B01, CMS-CSST-2021-B11. T. Z. is supported in part by the National Key Research and Development Program of China under Grant No.\,2020YFC2201503, the National Natural Science Foundation of China under Grant No.\,12275238 and No.\,11675143, the Zhejiang Provincial Natural Science Foundation of China under Grant No.\,LR21A050001 and LY20A050002,  and the Fundamental Research Funds for the Provincial Universities of Zhejiang in China under Grant No.\,RF-A2019015. T. L. is supported by the National Natural Science Foundation of China (NSFC) Grant No.\,12003008 and the China
Postdoctoral Science Foundation Grant No.\,2020M682393.
\end{acknowledgments}

\appendix

\section{\label{appendix-hadamada}Hadamard Regularization}
Let us consider the class of functions $F$ depending on the field point $\mathbf{x}$ as well as on two source points $\mathbf{z}_{1}$ and $\mathbf{z}_{2}$, and admitting, when the field point approaches one of the source points ($r_1\rightarrow|\mathbf{x}-\mathbf{z}_{1}|\rightarrow0$ for instance), an expansion of the type
\begin{equation}
\label{Fx-def}
F(\mathbf{x})=\sum_{-k_0\leqslant k\leqslant0}
r_{1}^{k}f_{k}(\mathbf{x};\mathbf{z}_{1},\mathbf{z}_{2}).
\end{equation}
We define the value of the function $F$ at the location $1$ to be the so-called Hadamard finite part, which is the average, with respect to the direction $\hat{\bm{n}}_{1}$, of the approach to point $1$, of the term with lowest-order coefficient in Eq.\,(\ref{Fx-def}),
\begin{equation}
F_{1}\equiv\frac{1}{4\pi}\int f_{0}(\mathbf{z}_1;\mathbf{z}_{1},\mathbf{z}_{2})
d\Omega_1,
\end{equation}
with $d\Omega_{1}$ a solid angle element centered on the location of the $1$-st object.

To derive the EOM, we need to give the Hadamard-regularized Newtonian potential $\bar{U}$, dCS scalar field $\vartheta$, and dCS-modified potential $\delta U$. These function can be rewritten via Eq.\,(\ref{Fx-def}) at the location of $1$-st object,
\begin{subequations}
\begin{align}
\bar{U}=\sum_{-1\leqslant k\leqslant0}
r_{1}^k\bar{u}_{k}(\mathbf{x};\mathbf{z}_{1},\mathbf{z}_{2}),\quad&\text{with}\quad\bar{u}_{-1}=m_1\quad\text{and}\quad
\bar{u}_0=\frac{m_2}{r_2}
=\frac{m_1}{|\mathbf{x}-\mathbf{z}_2|},\\
\vartheta=\sum_{-2\leqslant k\leqslant0}
r_{1}^k\tilde{\vartheta}_{k}(\mathbf{x};\mathbf{z}_{1},\mathbf{z}_{2}),\quad&\text{with}\quad
\tilde{\vartheta}_0=-\frac{\bm{n}\cdot\bm{\mu}_{2}}{r_2^2}=-\frac{\bm{n}\cdot\bm{\mu}_{2}}{|\mathbf{x}-\mathbf{z}_{2}|^2},\\
\delta U=\sum_{-3\leqslant k\leqslant0}r_1^k
\delta u_{k}(\mathbf{x};\mathbf{z}_{1},\mathbf{z}_{2})
\quad&\text{with}\quad
\delta u_{0}
=\frac{603}{3584}\frac{\zeta_{2}}{m_2^4}
\left(\frac{m_2}{r_2}\right)^3
\left[(\hat{\bm{n}}_{2}\cdot\bm{S}_{2})^2
-\frac{1}{3}(\bm{S}_{2}\cdot\bm{S}_{2})\right].
\end{align}
\end{subequations}
Such that the finite values of the Newtonian potential, scalar field, and dCS potential at location $1$ are
\begin{subequations}
\label{Hadamard-regularized-potential}
\begin{align}
\label{Hadamard-regularized-Newtonian-potential}
\bar{U}_1&=\frac{1}{4\pi}\int\bar{u}_{0}(\mathbf{z}_{1};\mathbf{z}_{1},\mathbf{z}_{2})d\Omega_1
=\frac{m_1}{|\mathbf{z}_{1}-\mathbf{z}_{2}|}
=\frac{m_1}{r},\\
\label{Hadamard-regularized-scalar}
\vartheta_1&=\frac{1}{4\pi}\int\tilde{\vartheta}_{0}(\mathbf{z}_{1};\mathbf{z}_{1},\mathbf{z}_{2})d\Omega_1
=\frac{\bm{n}\cdot\bm{\mu}_{2}}{|\mathbf{z}_{1}-\mathbf{z}_{2}|^2}
=\frac{\bm{n}\cdot\bm{\mu}_{2}}{r^2},\\
\label{Hadamard-regularized-dCS-potential}
\delta{U}_1&=\frac{1}{4\pi}\int\delta{u}_{0}(\mathbf{z}_{1};\mathbf{z}_{1},\mathbf{z}_{2})d\Omega_1
=\frac{603}{3584}\frac{\zeta_{2}}{m_2^4}
\left(\frac{m_2}{r}\right)^3
\left[(\hat{\bm{n}}\cdot\bm{S}_{2})^2
-\frac{1}{3}(\bm{S}_{2}\cdot\bm{S}_{2})\right],
\end{align}
\end{subequations}
respectively. The results for the location $2$, $\bar{U}_{2}$, $\vartheta_{2}$, and $\delta U_2$, are similar to Eq.\,(\ref{Hadamard-regularized-potential}). One can find that the finite part of Newtonian potential at location $1$ is just the value of Newtonian potential generated by the $2$-nd object at the $1$-st position, which is equivalent to omitting the self field. These finite results are used to derive the EOM of a black hole in the binary system (see Eq.\,(\ref{U2-theta2-deltaU2})), which is the same as that given in Ref.\,\cite{Yagi2012gw,Loutrel2018}, where the self-field contributions are omitted directly.

\section{\label{appendix-scalar}Estimation to the PN Order of Scalar Radiation}
In this appendix, we plan to analyze the PN order of scalar radiation. 

The multipole scalar radiations from non-compact sources term are estimated as
\begin{subequations}
\label{theta-sigma1}
\begin{align}
\label{theta-sigma1-mon}
\vartheta^{(\sigma_{1})}_{\rm mon}
&=-\frac{1}{R}\frac{1}{2\pi}
\int U(\partial_{i}\partial_{i}
\vartheta^{\rm(B)})d^3\mathbf{x}
\propto\frac{m}{R}\frac{\mu}{r^2}
\sim\mathcal{O}(1/c^6),\\
\label{theta-sigma1-dip}
\vartheta^{(\sigma_{1})}_{\rm dip}
&=-\frac{1}{R}\frac{1}{2\pi}\frac{\partial}{\partial t}
\int U(\partial_{i}\partial_{i}\vartheta^{\rm(B)})
(\hat{\mathbf{N}}\cdot\mathbf{x})d^3\mathbf{x}
\propto\frac{m}{R}\frac{\mu}{r^2}v
\sim\mathcal{O}(1/c^7),\\
\label{theta-sigma1-quad}
\vartheta^{(\sigma_{1})}_{\rm quad}
&=-\frac{1}{R}\frac{1}{4\pi}\frac{\partial^2}{\partial t^2}
\int U(\partial_{i}\partial_{i}\vartheta^{\rm(B)})
(\hat{\mathbf{N}}\cdot\mathbf{x})^2d^3\mathbf{x}
\propto\frac{m}{R}\frac{\mu}{r^2}v^2
\sim\mathcal{O}(1/c^8),
\end{align}
\end{subequations}
and
\begin{subequations}
\begin{align}
\label{theta-sigma2-mon}
&\vartheta^{(\sigma_{2})}_{\rm mon}
=\frac{2}{R}\frac{1}{\pi}\frac{\alpha}{\beta}\int\epsilon^{ijk}
(\partial_{i}\partial_{m}U)(\partial_{j}\partial_{m}V_{k})d^3\mathbf{x}
=\frac{2}{R}\frac{1}{\pi}
\epsilon^{ijk}\frac{\alpha}{\beta}
\oint\hat{N}^{i}\left[(\partial_{m}U)(\partial_{j}\partial_{m}V_{k})\right]R^2d\Omega
\propto\frac{1}{R^3}\rightarrow0,\\
\label{theta-sigma2-dip}
&\begin{aligned}
\vartheta^{(\sigma_{2})}_{\rm dip}
&=\frac{2}{R}\frac{1}{\pi}\frac{\alpha}{\beta}
\epsilon^{ijk}\frac{\partial}{\partial t}
\int(\partial_{i}\partial_{m}U)(\partial_{j}\partial_{m}V_{k})
(\hat{\mathbf{N}}\cdot\mathbf{x})d^3\mathbf{x}\\
&=\frac{2}{R}\frac{1}{\pi}\frac{\alpha}{\beta}
\epsilon^{ijk}\frac{\partial}{\partial t}
\left\{\oint\hat{N}_{j}[(\partial_{i}\partial_{m}U)(\partial_{m}V_{k})
(\hat{\mathbf{N}}\cdot\mathbf{x})]R^2d\Omega
\right\}
\propto\frac{1}{R^2}\rightarrow0,
\end{aligned}\\
\label{theta-sigma2-quad}
&\vartheta^{(\sigma_{2})}_{\rm quad}
=\frac{1}{R}\frac{1}{\pi}\frac{\alpha}{\beta}\epsilon^{ijk}
\frac{\partial^2}{\partial t^2}
\int(\partial_{i}\partial_{m}U)(\partial_{j}\partial_{m}V_{k})
(\hat{\mathbf{N}}\cdot\mathbf{x})^{2}d^3\mathbf{x}
\propto\frac{m}{R}\frac{m}{r^3}v^3
\sim\mathcal{O}(1/c^{11}).
\end{align}
\end{subequations}
The multipole scalar radiations from compact sources $\rho_1$ are estimated as
\begin{subequations}
\begin{align}
\label{theta-rho1-mon}
\vartheta^{(\rho_1)}_{\rm mon}
&=-\frac{1}{R}\sum_{A}\mu_{A}^{i}
\int\partial_{i}\delta^{(3)}(\mathbf{x}-\mathbf{z}_{A}(t_r))d^3\mathbf{x}=0,\\
\label{theta-rho1-dip}
\vartheta^{(\rho_1)}_{\rm dip}
&=-\frac{1}{R}\frac{\partial}{\partial t}
\sum_{A}\mu_{A}^{i}\int
[\partial_{i}\delta^{(3)}
(\mathbf{x}-\mathbf{z}_{A}(t_r))]
(\hat{\mathbf{N}}\cdot\mathbf{x})d^3\mathbf{x}
=\frac{1}{R}\hat{N}^{j}\frac{\partial}{\partial t}
\sum_{A}\mu_{A}^{i}\delta_{ij}=0,\\
\label{theta-rho1-quad}
\vartheta^{(\rho_1)}_{\rm quad}
&=-\frac{1}{R}\frac{1}{2}\frac{\partial^2}{\partial t^2}
\sum_{A}\mu_{A}^{i}
\int[\partial_{i}\delta^{(3)}(\mathbf{x}-\mathbf{z}_{A}(t_r))]
(\hat{\mathbf{N}}\cdot\mathbf{x})^{2}d^3\mathbf{x}
\propto\frac{m}{R}\frac{\mu}{r^2}
\sim\mathcal{O}(1/c^6).
\end{align}
\end{subequations}
The multipole scalar radiations from $\rho_2$ are estimated as
\begin{subequations}
\begin{align}
\label{theta-rho2-mon}
\vartheta^{(\rho_2)}_{\rm mon}
&=-\frac{1}{R}\sum_{A}\mu_{A}^{i}\dot{v}^{i}_{A}
\int\delta^{(3)}(\mathbf{x}-\mathbf{z}_{A}(t_{r}))d^3\mathbf{x}
\propto\frac{m}{R}\frac{\mu}{r^2}
\sim\mathcal{O}(1/c^{6}),\\
\label{theta-rho2-dip}
\vartheta^{(\rho_2)}_{\rm dip}
&=-\frac{1}{R}\frac{\partial}{\partial t}
\sum_{A}\mu_{A}^{i}\dot{v}^{i}_{A}\int
\delta^{(3)}(\mathbf{x}-\mathbf{z}_{A}(t_r))
(\hat{\mathbf{N}}\cdot\mathbf{z})d^3\mathbf{x}
\propto\frac{m}{R}\frac{\mu}{r^2}v
\sim\mathcal{O}(1/c^{7}),\\
\label{theta-rho2-quad}
\vartheta^{(\rho_2)}_{\rm quad}
&=-\frac{1}{R}\frac{1}{2}\frac{\partial^2}{\partial t^2}
\sum_{A}\mu_{A}^{i}\dot{v}^{i}_{A}
\int\delta^{(3)}(\mathbf{x}-\mathbf{z}_{A}(t_r))
(\hat{\mathbf{N}}\cdot\mathbf{x})^{2}d^3\mathbf{x}
\propto\frac{m}{R}\frac{\mu}{r^2}v^2
\sim\mathcal{O}(1/c^{8}).
\end{align}
\end{subequations}
The multipole scalar radiations from $\rho_3$ are estimated as
\begin{subequations}
\begin{align}
\label{theta-rho3-mon}
\vartheta^{(\rho_3)}_{\rm mon}
&=\frac{1}{R}\sum_{A}\mu_{A}^{i}v_{A}^{i}v_{A}^{j}
\int\partial_{j}\delta^{(3)}(\mathbf{x}-\mathbf{z}_{A}(t_r))d^3\mathbf{x}=0,\\
\label{theta-rho3-dip}
\vartheta^{(\rho_3)}_{\rm dip}
&=\frac{1}{R}\frac{\partial}{\partial t}
\sum_{A}\mu_{A}^{i}v_{A}^{i}v_{A}^{j}
\int\left[\partial_{j}\delta^{(3)}(\mathbf{x}-\mathbf{z}_{A}(t_r))\right]
(\hat{\mathbf{N}}\cdot\mathbf{x})d^3\mathbf{x}
\propto\frac{m}{R}\frac{\mu}{r^2}v
\sim\mathcal{O}(1/c^{7}),\\
\label{theta-rho3-quad}
\vartheta^{(\rho_3)}_{\rm quad}
&=\frac{1}{R}\frac{1}{2}\frac{\partial^2}{\partial t^2}
\sum_{A}\mu_{A}^{i}v_{A}^{i}v_{A}^{j}
\int\left[\partial_{j}\delta^{(3)}(\mathbf{x}-\mathbf{z}_{A}(t_r))\right]
(\hat{\mathbf{N}}\cdot\mathbf{x})^{2}d^3\mathbf{x}
\propto\frac{m}{R}\frac{\mu}{r^2}v^{2}
\sim\mathcal{O}(1/c^{8}).
\end{align}
\end{subequations}
Combining all the above results, we find the leading-order scalar radiations are given by $\vartheta_{\rm mon}^{(\sigma_{1})}$, $\vartheta_{\rm mon}^{(\rho_2)}$, and $\vartheta_{\rm quad}^{(\rho_1)}$. All dipole radiation is either strictly zero or higher than the order we consider.

\section{\label{appendix-multipole}Waveform from Source Terms}
The GW waveforms have given in subsection \ref{subsec:tensorradiation}. In this appendix, we plan to derive the same results through the method provided by Refs.\,\cite{Yagi2012pn,Yagi2016e}. The monopole formula of metric-deformation radiation is
\begin{equation}
\label{tensor-multipole}
\bar{k}_{ij}=\frac{4}{R}\int K_{ij}d^3\mathbf{x}.
\end{equation}
The source term $K_{ij}$ is given in Eq.\,(\ref{tensor-source}). The spatial component of the source of tensor perturbation is
\begin{equation}
\label{Kij}
K_{ij}=kT_{ij}^{(m)}+(1+h)\delta T_{ij}^{(m)}
+T_{ij}^{(\vartheta)}+\frac{1}{16\pi}\tilde{\Lambda}_{ij}
-2\alpha\tilde{C}_{ij}.
\end{equation}
The first and second terms in Eq.\,(\ref{Kij}) are of order $v^2S^2$, where $v$ represents the velocity and $S$ represents the spin angular momentum of binary BHs. Therefore, the contribution from these two terms will enter higher PN order. The third term comes from the EMT of the scalar field,
\begin{equation}
\label{Tij-theta}
T_{ij}^{(\vartheta)}=\beta\left[(\partial_{i}\vartheta)(\partial_{j}\vartheta)
-\frac{1}{2}\delta_{ij}(\partial_{k}\vartheta)(\partial_{k}\vartheta)\right],
\end{equation}
which will produce the part of SS (DD) coupling. The fourth term,
\begin{equation}
\label{Lambda-ij}
\begin{aligned}
\frac{1}{16\pi}\tilde{\Lambda}_{ij}
=\frac{1}{4\pi}\left[
-\delta_{ij}(\partial_{k}\bar{U})(\partial_{k}\delta U)
+(\partial_{i}\bar{U})(\partial_{j}\delta U)
+(\partial_{i}\delta U)(\partial_{j}\bar{U})\right],
\end{aligned}
\end{equation}
provides the MQ effect in gravitational waveforms. The last term is
\begin{equation}
\label{C-ij}
-2\alpha\tilde{C}_{ij}
=-4\alpha\epsilon_{r(i|m}\partial_{m}(\Box^2_{\eta}\hat{V}_{j)})
(\partial_{r}\vartheta)
+8\alpha\epsilon_{r(j|m}
(\partial_{m}\partial_{[i)}\hat{V}_{k]})
(\partial_{k}\partial_{r}\vartheta),
\end{equation}
which implies the parity-violating effect of GWs in dCS theory. Later, we will prove that the contribution of this term to GWs is zero. This shows again that GWs will not show the parity-violating effect without considering the cosmological expansion and background scalar field.

According to Eq.\,(\ref{tensor-multipole}), the contribution from Eq.\,(\ref{Tij-theta}) is
\begin{equation}
\label{radiation-Tij-theta}
\begin{aligned}
\frac{4}{R}\int T_{ij}^{(\vartheta)}d^3\mathbf{x}
&=\frac{4\beta}{R}\int
\left[(\partial_{i}\vartheta_1)(\partial_{j}\vartheta_2)
-\frac{1}{2}\delta_{ij}(\partial_{k}\vartheta_1)(\partial_{k}\vartheta_2)\right]
d^3\mathbf{x}
+(1\leftrightarrow2)\\
&=\frac{4\beta}{R}
\mu_{1}^{k}\mu_{2}^{l}
\int\left[\partial_{i}\partial_{k}\left(\frac{1}{r_{1}}\right)
\partial_{j}\partial_{l}\left(\frac{1}{r_{2}}\right)
-\frac{1}{2}\delta_{ij}
\partial_{q}\partial_{k}\left(\frac{1}{r_{1}}\right)
\partial_{q}\partial_{l}\left(\frac{1}{r_{2}}\right)\right]
d^3\mathbf{x}+(1\leftrightarrow2)\\
&=\frac{4\beta}{R}
\mu_{1}^{k}\mu_{2}^{l}
\Bigg[\partial^{(1)}_{ik}\partial^{(2)}_{jl}
\int\frac{1}{r_{1}}\frac{1}{r_{2}}d^3\mathbf{x}
-\frac{1}{2}\delta_{ij}\partial^{(1)}_{qk}\partial^{(2)}_{ql}
\int\frac{1}{r_{1}}\frac{1}{r_{2}}d^3\mathbf{x}\Bigg]
+(1\leftrightarrow2)\\
&=-\frac{8\pi\beta}{R}
\mu_{1}^{k}\mu_{2}^{l}
\left[\partial^{(1)}_{ik}\partial^{(2)}_{jl}r
-\frac{1}{2}\delta_{ij}\partial^{(1)}_{qk}\partial^{(2)}_{ql}r\right]
+(1\leftrightarrow2)\\
&=-\frac{4\mu}{R}\frac{75}{256}\zeta\gamma^3\frac{1}{\nu}
\left(\frac{\bm{S}_{1}}{m_1^2}\cdot\frac{\bm{S}_{2}}{m_2^2}\right)
\hat{n}^{i}\hat{n}^{j},
\end{aligned}
\end{equation}
and from Eq.\,(\ref{Lambda-ij}) is
\begin{equation}
\label{radiation-Lambda-ij}
\begin{aligned}
\frac{4}{R}
\int\frac{\tilde{\Lambda}_{ij}}{16\pi}d^3\mathbf{x}
&=\frac{1}{R}\frac{1}{\pi}
\int\left[(\partial_{i}\bar{U}_{1})(\partial_{j}\delta U_{2})
+(\partial_{j}\bar{U}_{1})(\partial_{i}\delta U_{2})
-\delta_{ij}(\partial_{k}\bar{U}_{1})(\partial_{k}\delta U_{2})\right]
d^3\mathbf{x}+(1\leftrightarrow2)\\
&=\frac{1}{R}\frac{1}{\pi}\delta C_Q\frac{m_1}{m_2}S_2^kS_2^l
\int\left[\partial_{k}\partial_{(i}\left(\frac{1}{r_{1}}\right)
\partial_{j)}\partial_{l}\left(\frac{1}{r_{2}}\right)
-\frac{1}{2}\delta_{ij}\partial_{q}\partial_{k}\left(\frac{1}{r_{1}}\right)
\partial_{q}\partial_{l}\left(\frac{1}{r_{2}}\right)
\right]d^3\mathbf{x}+(1\leftrightarrow2)\\
&=\frac{1}{R}\frac{1}{\pi}\delta C_Q\frac{m_1}{m_2}S_2^kS_2^l
\left[\partial^{(1)}_{ik}\partial^{(2)}_{jl}
\int\frac{1}{r_{1}}\frac{1}{r_{2}}d^3\mathbf{x}
-\frac{1}{2}\delta_{ij}\partial^{(1)}_{kq}\partial^{(2)}_{lq}
\int\frac{1}{r_{1}}\frac{1}{r_{2}}d^3\mathbf{x}\right]+(1\leftrightarrow2)\\
&=-\frac{2}{R}\delta C_Q\frac{m_1}{m_2}S_2^kS_2^l
\left[\partial^{(1)}_{ik}\partial^{(2)}_{jl}r
-\frac{1}{2}\delta_{ij}\partial^{(1)}_{qk}\partial^{(2)}_{ql}r\right]
+(1\leftrightarrow2)\\
&=\frac{4\mu}{R}\gamma^3\frac{603}{3584}\zeta
\left[\frac{m^2}{m_1^2}
\left(\frac{\bm{S}_1}{m_1^2}\right)^2
+\frac{m^2}{m_2^2}
\left(\frac{\bm{S}_2}{m_2^2}\right)^2\right]
\hat{n}^{i}\hat{n}^{j}.
\end{aligned}
\end{equation}
$(1\leftrightarrow2)$ represents the exchange of labels $1$ and $2$. The operators $\partial_{k}^{(1)}$ and $\partial_{k}^{(2)}$ are the derivative with respect to the coordinates of objects, $\mathbf{z}_{1}$ and $\mathbf{z}_{2}$. Here we only consider the contribution of cross-interaction terms. 

The GWs generated from Eq.\,(\ref{C-ij}) is
\begin{equation}
\label{radiation-Cij}
\frac{4}{R}\int(-2\alpha\tilde{C}_{ij})d^3\mathbf{x}
=\frac{8\alpha}{R}\int
[\epsilon_{rjm}
(\partial_{r}\partial_{k}\partial_{m}\partial_{i}\bar{V}_{k})
+\epsilon_{rim}
(\partial_{r}\partial_{k}\partial_{m}\partial_{j}\bar{V}_{k})]
\vartheta^{\rm(B)}d^3\mathbf{x}=0.
\end{equation}
Combining Eq.\,(\ref{radiation-Tij-theta}) and (\ref{radiation-Lambda-ij}), we obtain the gravitational waveforms of the metric deformation,
\begin{equation}
\bar{k}_{ij}
=\frac{4\mu}{R}\gamma^3\cdot\zeta
\left\{\frac{603}{3584}
\left[\frac{m^2}{m_1^2}
\left(\frac{\bm{S}_1}{m_1^2}\right)^2
+\frac{m^2}{m_2^2}
\left(\frac{\bm{S}_2}{m_2^2}\right)^2\right]
-\frac{75}{256}\frac{1}{\nu}
\left(\frac{\bm{S}_{1}}{m_1^2}\cdot\frac{\bm{S}_{2}}{m_2^2}\right)
\right\}\hat{n}^{i}\hat{n}^{j}
=-\frac{4\mu\gamma^3}{R}
\delta\varpi\cdot\hat{n}^{i}\hat{n}^{j},
\end{equation}
which is consistent with the result given in Eq.\,(\ref{kij}).

\bibliographystyle{apsrev4-2}
\bibliography{reference}
\end{document}